\documentclass[]{emulateapj}
\usepackage{apjfonts}
\newcommand{\be}{\begin{equation}}
\newcommand{\ee}{\end{equation}}
\newcommand{\bea}{\begin{eqnarray}}
\newcommand{\eea}{\end{eqnarray}}

\newcommand{\Msun}{M_{\odot}}

\shortauthors{CONROY, GUNN, AND WHITE}
\shorttitle{THE PROPAGATION OF UNCERTAINTIES IN SPS MODELING}
\begin{document}
\journalinfo{The Astrophysical Journal}
\submitted{Submitted to the Astrophysical Journal}

\title{The Propagation of Uncertainties in stellar population
  synthesis modeling I: The relevance of uncertain aspects of stellar
  evolution and the IMF to the derived physical properties of
  galaxies}
\author{Charlie Conroy} 
\author{James E. Gunn} 
\affil{Department of Astrophysical Sciences, Princeton University,
  Princeton, NJ 08544, USA}
\author{Martin White}
\affil{Departments of Physics and Astronomy, 601 Campbell Hall,
University of California Berkeley, CA 94720, USA}

\begin{abstract}

  The stellar masses, mean ages, metallicities, and star formation
  histories of galaxies are now commonly estimated via stellar
  population synthesis (SPS) techniques.  SPS relies on stellar
  evolution calculations from the main sequence to stellar death,
  stellar spectral libraries, phenomenological dust models, and
  stellar initial mass functions (IMFs) to translate the evolution of
  a multi-metallicity, multi-age set of stars into a prediction for
  the time-evolution of the integrated light from that set of stars.
  Each of these necessary inputs carries significant uncertainties
  that have until now received little systematic attention.  The
  present work is the first in a series that explores the impact of
  uncertainties in key phases of stellar evolution and the IMF on the
  derived physical properties of galaxies and the expected luminosity
  evolution for a passively evolving set of stars.  A Monte-Carlo
  Markov-Chain approach is taken to fit near-UV through near-IR
  photometry of a representative sample of low- and high-redshift
  galaxies with this new SPS model.  Significant results include the
  following: 1) including uncertainties in stellar evolution, stellar
  masses at $z\sim0$ carry errors of $\sim0.3$ dex at 95\% CL with
  little dependence on luminosity or color, while at $z\sim2$, the
  masses of bright red galaxies are uncertain at the $\sim0.6$ dex
  level; 2) either current stellar evolution models, current
  observational stellar libraries, or both, do not adequately
  characterize the metallicity-dependence of the thermally-pulsating
  asymptotic giant branch phase; 3) conservative estimates on the
  uncertainty of the slope of the IMF in the solar neighborhood imply
  that luminosity evolution per unit redshift is uncertain at the
  $\sim0.4$ mag level in the $K-$band, which is a substantial source
  of uncertainty for interpreting the evolution of galaxy populations
  across time.  Any possible evolution in the IMF, as suggested by
  several independent lines of evidence, will only exacerbate this
  problem.  4) Assuming a distribution of stellar metallicities within
  a galaxy, rather than a fixed value as is usually assumed, can yield
  important differences when considering bands blueward of $V$, but is
  not a concern for redder bands.  Spectroscopic information may
  alleviate some of these concerns, though uncertainties in the
  stellar spectral libraries and the importance of non-solar abundance
  ratios have not yet been systematically investigated in the SPS
  context.

\end{abstract}



\section{Introduction}
\label{s:intro}

The spatially-integrated UV, optical, and near-IR light of a galaxy
contains a wealth of information regarding its physical properties.
The spectrum of this light is governed principly by the star formation
and metal enrichment histories of a galaxy in conjunction with stellar
evolution and attenuation by interstellar dust.  Combining these
ingredients in order to predict the spectrum of a galaxy is known as
stellar population synthesis (SPS), and has an extensive history
\citep[e.g.][]{Tinsley76, Tinsley80, Bruzual83, Renzini86, Bruzual93,
  Worthey94, Maraston98, Leitherer95, Fioc97, Vazdekis99, Yi03a,
  Bruzual03, Jimenez04, Maraston05, CidFernandes05, Ocvirk06}.

Sophisticated SPS models have allowed for an enormous body of work
aimed at constraining physical parameters of galaxies by comparing the
observed spectroscopic and/or photometric properties of $z\sim0$
galaxies to SPS models.  Physical parameters that are constrained by
SPS include, but are not limited to, stellar masses
\citep[e.g.][]{Bell03, Kauffmann03a, Panter07}, star formation
histories and rates \citep[e.g.][]{Kauffmann03b, Panter07}, and
metallicities \citep[e.g.][]{Gallazzi05}.  SPS is now routinely used
not only at $z\sim0$ but also at higher redshifts in order to
interpret the observed broad-band colors and spectra of galaxies
\citep[e.g.][]{Shapley05, Daddi07a, Kriek06a, Bundy06}.

Key aspects of stellar evolution theory --- an essential ingredient of
any SPS model --- have been extensively tested against observations
including globular clusters and star counts in the solar neighborhood,
stellar halo, and the Large and Small Magellenic Clouds (LMC and SMC,
respectively), suggesting that in many respects they are sufficiently
reliable to be applied in SPS modeling of observed galaxies.  However,
there are significant phases in stellar evolution that are still not
well understood, both observationally and theoretically.  Such phases
include blue stragglers (BSs), the horizontal branch (HB), the
asymptotic giant branch (AGB), and thermally pulsating AGBs (TP-AGBs).
Relatedly, the different evolutionary path(s) taken by binary systems,
which appear to constitute the majority of all stellar systems, is not
well-understood.  These phases are particularly important for SPS
models because stars in such phases are more luminous than the main
sequence (in some cases by several orders of magnitude), and in the
case of AGBs they can dominate the red/IR light in some age ranges of
a coeval set of stars, while in the case of the HB and BS they can
contribute importantly to the blue and UV light.  These uncertainties
will thus impact our ability to derive physical properties of galaxies
\citep[e.g.][]{Charlot96a, Charlot96b, Yi03, LeeHC07}.

The potential importance of uncertain phases of stellar evolution was
highlighted recently by the different treatments of TP-AGBs in the SPS
models of \citet{Maraston05} and \citet{Bruzual03}.  These models,
when applied to high-redshift galaxies where light is dominated by
$\sim0.2-2$ Gyr old populations, result in {\it systematic} factors of
$\sim2$ differences in mass and age \citep{Maraston06, Bruzual07,
  Kannappan07}.  This example demonstrates the need for understanding
in detail how the uncertainties in stellar evolution propagate into
estimates of physical properties of galaxies.  

In addition to uncertainties in stellar evolution, there are also
substantial uncertainties in our knowledge of the stellar initial mass
function \citep[IMF][]{Kroupa01}.  The IMF effects not only the
normalization of the mass-to-light ratio but also, to a lesser degree,
the colors of a coeval set of stars \citep{Tinsley80}.  The IMF thus
plays an important role in deriving physical properties of galaxies
\citep[e.g.][]{Papovich01, LeeHC04a}.  The IMF also controls the rate
of luminosity evolution for a passively evolving system, and is thus
of fundamental importance for interpreting the evolution of galaxy
populations across time \citep[e.g.][]{Tinsley80, Yi03}.

In this paper we present a new SPS code that was developed in order to
address the impact of uncertainties in stellar structure and evolution
on the derived physical properties of galaxies.  This new SPS model is
then applied to a variety of galaxies drawn from the Sloan Digital Sky
Survey \citep[SDSS][]{York00} and the Two Micron All Sky Survey
\citep[2MASS][]{Jarrett00} at $z\sim0$, and from a sample of luminous
red galaxies at $z\sim2$, in order to investigate the effects of
uncertain phases of stellar evolution on derived physical properties
such as stellar mass and star formation history.  In addition to the
uncertain aspects of stellar evolution, we also explore the effects of
the IMF and the potential biases that arise when assuming a single
metallicity for all stars in a galaxy rather than the more physical
assumption of a distribution of stellar metallicities.

Where necessary, we adopt a flat $\Lambda$CDM cosmology with
$(\Omega_m, \Omega_\Lambda,h)=(0.26,0.74,0.72)$.  Throughout,
mass-to-light ratios are quoted in units of $M_\Sol/L_\Sol$.

\section{Methods: from physical properties to integrated light}
\label{s:methods}

\subsection{Overview}

This section describes the steps involved in SPS modeling.  Briefly,
the approach utilizes stellar evolution calculations from the zero-age
main sequence to stellar death to generate isochrones, which are the
age-dependent positions in the HR diagram of a coeval population of
stars.  A combination of empirical and empirically-calibrated
theoretical stellar libraries are then used to assign a full spectrum
to each point in the HR diagram.  The resulting integrated light from
a coeval set of stars (referred to as a `simple stellar population',
or SSP) is then simply a sum of all the spectra along an isochrone,
weighted by the number of stars at a given stellar mass, i.e. the IMF.
Finally, the integrated light from a galaxy with a complex star
formation history (SFH) is a convolution of a time-dependent SSP with
the SFH and a prescription for attenuation by dust.  These steps are
described in detail in the following sections.  The uncertainties
associated with SPS modeling are explored in depth in
$\S$\ref{s:uncert}.

\subsection{Stellar evolution}

The core of any SPS model is the stellar evolution tracks that allow
one to follow the evolution of stars of any mass from the zero-age
main sequence to later evolutionary stages.  The tracks must be
sufficiently well sampled in both mass and time so that isochrones can
be constructed.  Many groups have produced publically available
libraries of stellar evolution models \citep[e.g.][]{Bertelli94,
  Girardi00, Cioni06, Schaller92, Cassisi00, Yi01, Dotter07}.

We make use of the latest set of models from the Padova
group\footnote{The Padova isochrones are publically available on the
  web: \texttt{http://stev.oapd.inaf.it/cgi-bin/cmd}.}
\citep{Marigo07a, Marigo08}.  The library of isochrones includes
stars with initial masses $0.15\leq M\leq100 \,\Msun$ spanning ages
$10^{6.6}<t<10^{10.2} $ yrs, with outputs at equally spaced intervals
of $\Delta(\rm{log}(t/yr))=0.05$, and for metallicities in the range
$10^{-4}<Z<0.030$, in equally spaced intervals of ${\rm log}(Z)=0.1$.
In the following we adopt $Z_\Sun=0.019$.  These isochrones include a
detailed treatment of the TP-AGB phase that has been calibrated
against near-IR data from the LMC and SMC.  The Padova models are
supplemented in the mass range $0.10\leq M<0.15 \,\Msun$ with the
non-evolving stellar models of \citet{Baraffe98}.  Stars at this low
mass range neither evolve off the main sequence nor contribute
significantly to the total light emitted from a galaxy.  Nonetheless
they contribute substantially to the total stellar {\emph mass} of a
galaxy.

\subsection{Spectral libraries}
\label{s:stellib}

In addition to accurate stellar evolution models, the SPS formalism
also requires an extensive, well-calibrated spectral library.  We make
use of the semi-empirical BaSeL3.1 library \citep{Lejeune97,
  Lejeune98, Westera02}, which is a compilation of model atmosphere
calculations for a wide range in effective temperature and surface
gravity, for the full range of metallicities in the Padova stellar
evolution models.  The library spans the wavelength range
$91\AA-160\mu m$ at a resolving power of
$\lambda/\Delta\lambda\approx200-500$.  The library does not include
stars hotter than $50,000$ K.  We approximate the spectra of such
stars as pure blackbodies.

The original model atmosphere calculations have been re-calibrated by
coupling the libraries to theoretical isochrones and comparing to
globular cluster color-magnitude diagrams for a range in metallicity
\citep{Westera02}.  Corrections to the original atmospheric models are
significant especially for $M$ stars because the models do not include
the important effect of line blanketing due to molecular lines.  As
discussed in detail in \citet{Westera02}, the BaSeL library cannot
simultaneously match the color-magnitude globular cluster data and the
observed $UBVRIJHKL$ color-temperature relations for individual stars.
It is not known whether the color-temperature relations or the
theoretical isochrones are the source of the discrepancy.  This
implies that the stellar library itself suffers unknown systematic
uncertainties because of the choice to calibrate the models against
the globular cluster data.

The BaSeL3.1 stellar spectral library does not include the spectra of
TP-AGBs.  It is critical to include the spectra of such stars since
they dominate the light output of intermediate age stellar populations
\citep{Maraston05}.  We make use of average TP-AGB spectra compiled by
\citet{Lancon02} from more than 100 optical/near-IR spectra presented
in \citet{Lancon00}.  The library spans the wavelength range $0.5-2.5
\mu m$ and is separated into oxygen-rich and carbon-rich spectra. The
O-rich and C-rich spectra are further divided into nine and five bins
in $I-K$ color, respectively.  $I-K$ color is then converted into
$T_{\rm eff}$ with the theoretical metallicity-dependent
color-temperature relations in \citet{Bessell91}.

It is important to note that the metallicity of the observed spectra
in this library are not known.  \citet{Lancon02} assume that the stars
are all approximately solar metallicity because approximately three
quarters of the spectra are from the field of the Milky Way.  The
authors advocate extending the library to non-solar metallicities with
the following prescription: take $T_{\rm eff}$ and $Z$ from the
stellar evolution models, use the color--metallicity--temperature
relations of \citet{Bessell91} to infer the $I-K$ color, and then use
the library spectrum that most closely matches that color.  Moreover,
the spectra are not binned as a function of luminosity (or surface
gravity).  Rather, it is assumed that the spectra are not sensitive to
luminosity.

Both the empirically unknown luminosity and metallicity dependence of
the spectra introduce significant uncertainties that are difficult to
quantify.  Rather than introduce parameters that can be varied to
encompass the possible effects of these uncertainties, we take the
following approach.  The standard prescription mentioned above is
utilized to assign spectra to TP-AGB stars.  Any uncertainty in the
conversion between theoretical quantities ($L_{\rm bol}$ and $T_{\rm
  eff}$) and the emergent spectra is incorporated into uncertainties
in the theoretical quantities themselves, as discussed in
$\S$\ref{s:tpagb}.  We take this approach because uncertainties in the
theoretical quantities are highly degenerate with uncertainties in the
relation between theoretical quantities and emitted spectra, so it is
simpler to combine all uncertainties into the former.

Neither the spectral libraries nor the stellar evolution calculations
used herein account for the possibility of non-solar abundance ratios.
We return to this point in $\S$\ref{s:other}.

\subsection{The initial mass function}
\label{s:imf}

The initial distribution of stellar masses along the main sequence,
known as the stellar initial mass function or IMF, has been studied
extensively for decades \citep[e.g.][]{Salpeter55, Scalo86, Scalo98,
  Kroupa01, Chabrier03}.  Of particular relevance for SPS is the
logarithmic slope of the IMF, especially near $\sim1\Msun$ for old
stellar populations, and its possible universality (both in space and
time).  Direct measurements of the IMF in the Galaxy are extremely
challenging \citep[for extensive discussion of the difficulties
see][see also $\S$\ref{s:uimf}]{Scalo98, Kroupa01}.

In order to explore the importance of the IMF, we adopt the form
advocated by \citet{vanDokkum08}:
\noindent
\be
\label{eq:imf}
\Phi \equiv \frac{\rm{d}n}{\rm{dln}M} = \left\{
\begin{array}{c@{\quad}l}
A_l (0.5 n_c m_c)^{-x}
\exp\left[\frac{-(\log M - \log m_c)^2}{2\sigma^2}\right] & (M\leq n_c m_c) \\
A_h M^{-x} & (M>n_c m_c),
\end{array} \right.
\ee
\noindent
with $A_l = 0.140$, $n_c = 25$, $x=1.3$, $\sigma=0.69$, and
$A_h=0.158$. Variation of the IMF is incorporated in the
characteristic mass $m_c$.  Throughout we adopt lower and upper
mass cut-offs of $0.1\Msun$ and $100\Msun$, respectively.

Equation \ref{eq:imf} is almost identical to the form advocated by
\citet{Chabrier03} for $m_c=0.08$, and is also very similar to the
piece-wise power-law IMF proposed by \citet{Kroupa01}, again for
$m_c=0.08$.  From fitting the color and luminosity evolution of
cluster ellipticals \citet{vanDokkum08} suggests that $m_c\sim2$ at
$z\gtrsim 4$.  In $\S$\ref{s:uimf} we discuss further the existing
evidence for an IMF that evolves in time and discuss more generally
the importance of the IMF in SPS modeling.  Its importance in
understanding galaxy evolution was first discussed in
\citet{Tinsley80}; we discuss its importance in a modern context in
$\S$\ref{s:whyimf}.

It has been known since the work of \citet{Tinsley80} that the
logarithmic slope of the IMF at the main-sequence turn-off point
determines the rate of luminosity evolution of a passively evolving
system (for a system with ongoing star-formation it is the IMF in
addition to the star formation history that determines the rate of
luminosity evolution).  For this reason it is important that the
logarithmic slope of the IMF be continuous.  Thus, while the IMF of
\citet{Kroupa01}, which is a piece-wise power-law, is similar in
magnitude to other continuous IMFs, such as the one used herein, the
continuous variety must be preferred since they do not introduce
artificial jumps in the luminosity evolution of passive systems.  This
issue is especially acute since the \citet{Kroupa01} IMF changes
logarithmic slope at $1\Msun$, precisely the turn-off point for an old
stellar population.  We return to these issues in $\S$\ref{s:uimf} and
$\S$\ref{s:whyimf}.

\begin{figure}[t!]
\plotone{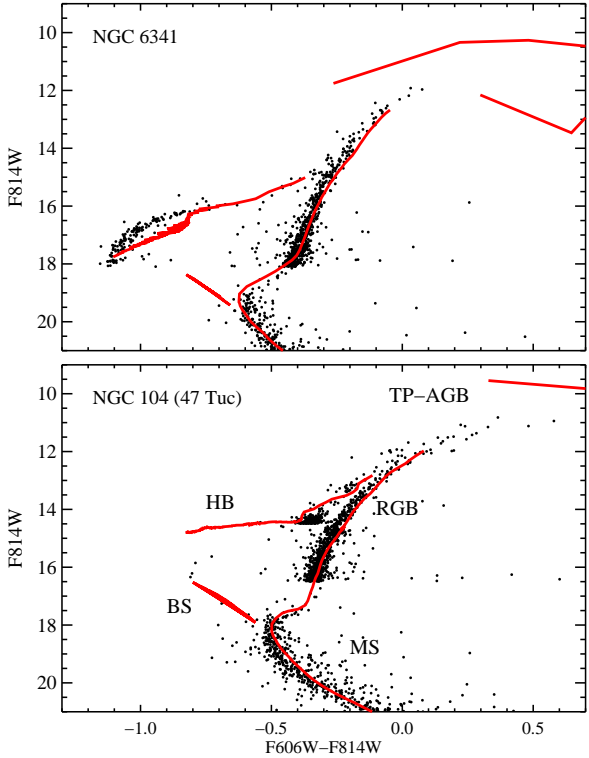}
\vspace{1.0cm}
\caption{Observed CMD of two globular clusters from the {\it HST}
  \citep{Brown05} compared to our SPS model.  The important
  evolutionary phases are labeled in the figure, including the main
  sequence (MS), blue stragglers (BS), the horizontal branch (HB), red
  giant branch (RGB), and the thermally-pulsating AGBs (TP-AGB).  The
  lines are not a fit to the points; they are however similar in
  metallicity and age to the observed globular clusters and thus
  should adequately represent the data.  In both panels only a
  fraction of the data is plotted below the RGB for clarity. {\it Top
    Panel:} NGC 6341 has an age of 14.5 Gyr and a metallicity of
  [Fe/H]$=-2.14$ \citep{Brown05}; the model has an age of 14.1 Gyr and
  a metallicity of [Fe/H]$=-2.28$.  Notice the clear blueward
  extension of the HB, as is expected for metal-poor globular
  clusters.  {\it Bottom Panel:} NGC 104 has an age of 12.5 Gyr and a
  metallicity of [Fe/H]$=-0.7$ \citep{Brown05}; the model has an age
  of 12.6 Gyr and a metallicity of [Fe/H]$=-0.68$.}
\label{fig:obshr}
\vspace{0.5cm}
\end{figure}

\subsection{SSPs}

With the isochrones, composite spectral library, and IMF in hand, we
are now in a position to construct SSPs, which are the building blocks
of SPS modeling.  The spectrum from a coeval set of stars of metallicity
$Z$ at time $t$ after birth can be described as:
\noindent
\be\label{eqn:ssp}
S(t,Z) = \int_{M_i^l}^{M_i^u(t)} \, \Phi(M_i)\Lambda[L(M_i,Z,t),T(M_i,Z,t),Z]\,{\rm d}M_i
\ee
\noindent
where $M_i$ is the initial (zero-age main sequence) mass, $M_i^l$ and
$M_i^u$ are lower and upper mass cut-offs, $\Phi$ is the IMF,
$\Lambda$ is a spectrum from the stellar library, and $L$ and $T$ are
the bolometric luminosity and effective temperature of a star of mass
$M_i$ and metallicity $Z$.  In the following we adopt
$M_i^l=0.1\Msun$.  $M_i^u(t)$, $L(t),$ and $T(t)$ are determined by
stellar evolution.

The mass of a coeval set of stars is:
\noindent
\be
M(t) = \int_{M_i^l}^{M_i^u(t)} \, \Phi(M_i) \, M_{\rm evol}(M_i) {\rm d}M_i \,+\, M_{\rm rem}
\ee
\noindent
where $M_{\rm evol}(M)$ is the evolved mass of a star of initial mass
$M_i$ and $M_{\rm rem}$ is the amount of mass locked up in remnants
including white dwarfs, neutron stars, and black holes.  

We follow the prescription of \citet{Renzini93} in assigning remnant
masses to dead stars: stars with initial masses $M_i\geq40\Msun$ leave
a remnant black hole of mass $0.5M_i$, initial masses $8.5\leq
M_i<40\Msun$ leave behind a $1.4\Msun$ neutron star, and initial
masses $M_i<8.5\Msun$ leave a white dwarf of mass $0.077M_i + 0.48$.
This prescription is standard in SPS \citep{Maraston98, Bruzual03},
although clearly these relations are not known to great precision
\citep[see][for a recent summary of the constraints on the
initial--final mass relation for white dwarfs]{Catalan08}.  For most
purposes the above prescription amounts to a normalizing factor in the
mass-to-light ratio, and it is thus of little concern for the present
work.  However, if these relations depend on other parameters such as
metallicity then they will have to be considered in greater detail.

\subsection{Isochrone synthesis}

The SSPs generated in the previous section describe the spectral
evolution of a coeval set of stars.  The final step in SPS modeling is
to generate the time-dependent spectrum of a galaxy made of stars of
varying ages.  This flux arises from the following integral:
\noindent
\be
F(t) = \int_0^t\, \Psi(t-t')S(t',Z)e^{-\hat{\tau}_\lambda(t')}\,{\rm d}t'
\ee
\noindent
where $\Psi$ is the star formation rate, $S_\nu(t,Z)$ is an SSP at
time $t$ and metallicity $Z$, and $e^{-\hat{\tau}_\lambda(t)}$
describes the attenuation of starlight by dust.  Note that we assume
that the metallicity is time-independent.  This is a common approach
in SPS modeling, though see \citet{Panter07} who allow $Z(t)$.

The star formation rate is assumed to commence at some epoch $T_{\rm
  start}$.  We allow for star formation to be characterized by two
components.  One component is a constant level of star formation
quantified as the fraction, $C$, of star formation occurring in this
mode.  The second component exponentially decreases with time and is
characterized by the e-folding timescale $\tau$.  Thus, for the star
formation rate we have:
\noindent
\be
\Psi(t) =  
 \frac{(1-C)}{\tau}\,\frac{e^{-t/\tau}}{e^{-T_{\rm start}/\tau}-e^{-T_{\rm univ}/\tau}} + \frac{C}{T_{\rm univ}-T_{\rm start}} \, , \,\,\,\, T_{\rm start} \leq t \leq T_{\rm univ}
\ee
\noindent
where $\Psi$ is normalized such that one solar mass of stars is
created over the age of the Universe, $T_{\rm univ}$.  Note that
mass-to-light ratios do not depend on the normalization of the star
formation rate since both the mass and light are integrals over
$\Psi$.  Our goal will be to determine the mass-to-light ratios of
galaxies and then to multiply by the measured galactic light to infer
the total stellar mass; our results will thus be entirely insensitive
to this normalization of $\Psi$.  

We adopt the simple two-component dust model of \citet{Charlot00},
where $\hat{\tau}(t)$ is given by
\noindent
\be
\hat{\tau}_\lambda(t) \equiv \left\{ \begin{array}{lc}
 \hat{\tau}_1(\lambda/5500\AA)^{-0.7} &\, t \leq 10^7 \,\,{\rm yr} \\
 \hat{\tau}_2(\lambda/5500\AA)^{-0.7} &\, t > 10^7 \,\,{\rm yr}
\end{array} \right.
\ee
\noindent
In this model, the optical depth for young stellar systems is
associated with dust in molecular clouds, in which the young stars are
embedded.  Following the disruption of molecular clouds, which occurs
on a timescale of $\sim10^7$ yr \citep{Blitz80}, the optical depth is
associated with a uniform screen across the galaxy.  \citet{Charlot00}
found that values of $\hat{\tau}_1\sim1.0$ and $\hat{\tau}_1\sim0.3$
adequately described an array of observational data.  In the following
these are left as free parameters.  The wavelength-dependence was also
constrained by the available data in \citet{Charlot00} although it too
could be left as a free parameter.  We leave it fixed for simplicity.

\begin{deluxetable}{lll}
\tablecaption{Summary of SPS Parameters}
\tablehead{ \colhead{Parameter} & \colhead{Description} &
\colhead{Range} }
\startdata\\
$\Delta_T$ & Shift in ${\rm log}(T_{\rm eff})$ along the TP-AGB & $[-0.2,0.2]$\\
$\Delta_L$ & Shift in ${\rm log}(L_{\rm bol})$ along the TP-AGB & $[-0.4,0.4]$ \\
$f_{BHB}$ & Fraction of blue HB stars & $[0.0,0.5]$ \\
$S_{BS}$ & Specific frequency of blue straggler stars & $[0,10]$ \\
$\tau$ & SFR e-folding time (Gyr) & $[0,\infty)$ \\
$C$ & Fraction of mass formed in a  & $[0,1]$\\
   &  constant mode of SF & \\
$T_{\rm start}$ & Age of Universe when SF commences (Gyr) & [$0.0,5.0]$\tablenotemark{a}\\
$\hat{\tau}_1$ & Extinction surrounding young stars & $[0,\infty)$ \\
$\hat{\tau}_2$ & Extinction surrounding old stars & $[0,\infty)$\\
$Z$ & Stellar metallicity & $[0.0001,0.030]$\\
$m_c$ & Characteristic mass of the IMF & $[0.08,2.0]$\\
\enddata
\tablenotetext{a}{The upper bound on the value of $T_{\rm start}$ is set to $T_{\rm univ}$ for galaxies at a redshift where the age of the Universe is younger than 5 Gyr.}
\label{t:models}
\end{deluxetable}

\section{Important Unknowns: stellar evolution, the IMF, and the
  distribution of stellar metallicities}
\label{s:uncert}

As discussed in the Introduction, there are numerous phases of stellar
evolution that are not well understood on both theoretical and
observational grounds.  In this section we discuss three phases of
particular importance: the TP-AGB, HB, and blue stragglers, and how
these phases can effect the integrated light from a galaxy.  These
phases are important in SPS due to their high bolometric luminosity
relative to the luminosity of main-sequence turn-off stars.  There are
additional phases that are not well understood, including post-AGB and
Wolf-Rayet stars, that we will not discuss because they are thought to
be important primarily in the UV and for very young populations,
respectively --- two observational regimes that we will not address
herein.  In this section we also explore the importance of the IMF and
the impact of assuming a distribution of stellar metallicities, rather
than a single metallicity, on the derived properties of galaxies.

In order to guide the eye, in Figure \ref{fig:obshr} we plot the
color-magnitude diagram (CMD) for two globular clusters observed with
the {\it Hubble Space Telescope} \citep[{\it HST};][]{Brown05}, and
compare them to stellar evolution tracks for models with ages and
metallicities comparable to the observed clusters.  This figure
highlights the locations of the various important phases of stellar
evolution that will be discussed in the following sections.

\begin{figure}[t!]
\plotone{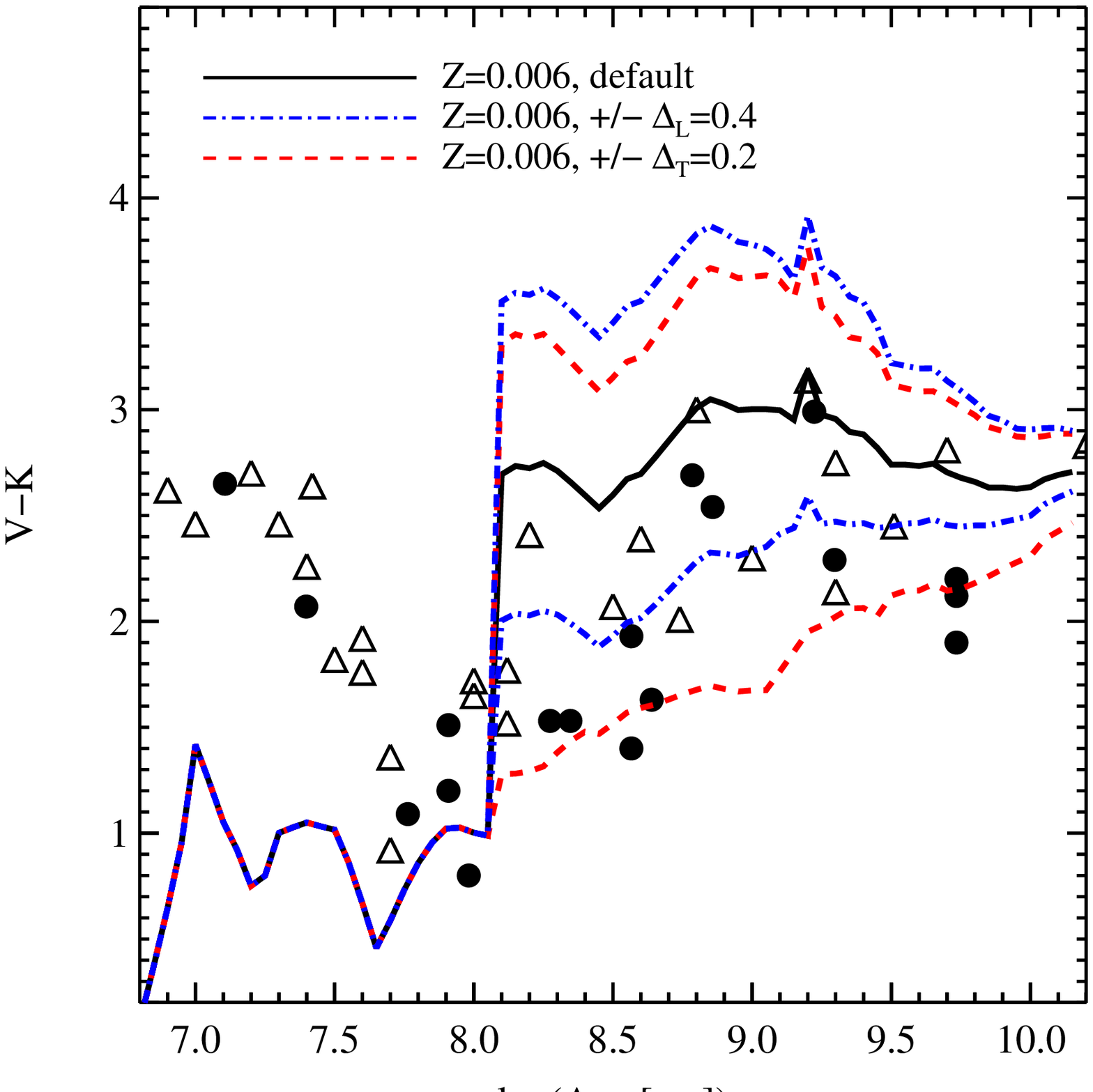}
\vspace{0.5cm}
\caption{$V-K$ colors of LMC star clusters as a function of cluster
  age.  Data are from the compilations of \citet[][{\it filled
    circles}]{Persson83} and \citet[][{\it open triangles}]{Kyeong03}.
  Ages for the \citet{Persson83} sample are adopted from
  \citet{Girardi95}.  The average metallicity of the LMC is
  $Z\approx0.006$ \citep{Cioni06}.  These data are compared to our
  default SPS model ({\it solid line}), and for a model with
  variations in the TP-AGB temperatures and luminosities by $\pm0.2$
  dex and $\pm 0.4$ dex ({\it dashed and dot-dashed lines},
  respectively).  Note that the standard stellar tracks without
  modification to the TP-AGB stars ({\it solid lines}) predict $V-K$
  colors redder than the majority of observed clusters.}
\label{fig:lmc_vk}
\vspace{0.5cm}
\end{figure}

\subsection{Thermally-pulsating AGB stars}
\label{s:tpagb}

Thermally-pulsating AGB stars (TP-AGBs) are an extremely difficult
phase of stellar evolution to model.  The thermal pulses are thought
to be driven by an instability arising when stars undergo helium shell
burning.  Hydrogen burning above the helium shell dumps helium ash
into the helium burning shell, which increases the energy output of
the helium shell.  Since the helium shell is both thin and partially
degenerate, the shell cannot readjust to the increased energy
production, and a thermal runaway ensues until the star can readjust.
This instability occurs in stars with initial masses $\lesssim5\Msun$
and hence for stellar populations with ages $\gtrsim10^8$ years
\citep{Marigo08}.

As mentioned in the Introduction, the importance of this phase of
stellar evolution recently became apparent when it was revealed that
the SPS codes of \citet{Bruzual03} and \citet{Maraston05} produce
systematically different predictions for the masses and ages of
intermediate age galaxies.  It quickly became clear that the different
treatment of TP-AGBs in these models, which dominate the light in the
near-IR for stellar populations several gigayears old, was the source
of the difference \citep{Maraston06}.

Progress in understanding the TP-AGB phase is hampered not only by the
aforementioned difficulties in theoretical modeling but also by the
lack of observations of stars in this phase.  Observations are
difficult because, while such stars are intrinsically luminous, they
are also very rare due to the short amount of time stars spend in this
phase.  Due to their rareness, they are often not seen in star
clusters, and thus the standard technique of comparing SSPs to
observed globular clusters does not shed light on the TP-AGB phase
(see e.g. Figure \ref{fig:obshr}).  The TP-AGB phase is best studied
in the field of the Milky Way and in the LMC and SMC --- regions where
bright stars can be resolved in large numbers by ground-based
telescopes.  Appealing to data collected from such large regions
introduces additional uncertainties, however, because the stars in
such a sample span a range of ages and metallicities, neither of which
are known.

The uncertainties associated with TP-AGB stars are modeled in the
following manner: the location of the TP-AGB track (see e.g. Figure
\ref{fig:obshr}) in the HR diagram is specified by two parameters,
$\Delta_L$ and $\Delta_{T}$, that characterize the shift in
${\rm log}(L_{\rm bol})$ and ${\rm log}(T_{\rm eff})$, respectively, with
respect to the default tracks provided by the models of
\citet{Marigo08}.  For simplicity, these shifts are independent of
time for an SSP.

This simple parameterization is motivated by the fact that several
uncertain steps are required in order to transform a theoretical
TP-AGB star into observable properties.  First, its luminosity and
temperature must be specified, as a function of metallicity, as given
by stellar evolution calculations.  Next, the luminosity, temperature,
and metallicity must be converted to a full spectrum for the star.
This step is made with observed spectra of a sample of TP-AGB stars
(with unknown metallicities) since theoretical TP-AGB spectra are
extremely difficult to calculate.  Finally, reddening by circumstellar
dust may be important, although our default models do not include this
effect.  Each of these steps carries significant uncertainties.  Since
we are only interested in transforming a theoretical TP-AGB star into
observable properties, we have incorporated all of these uncertainties
into the two parameters described above.

For example, if after fitting our model to an observed galaxy we find
non-zero values for either or both of these parameters, then we would
conclude that either our default stellar evolution model is
inaccurate, or our stellar spectral library is flawed, or there is
substantial circumstellar dust that we are not including, or some
combination of all three of these.

While the details of the TP-AGB phase, and hence the values of the
parameters $\Delta_L$ and $\Delta_{T}$ are uncertain, data from both
the Milky Way and Magellanic Clouds provide valuable constraints on
these uncertainties.  Figure \ref{fig:lmc_vk} plots the $V-K$ color of
star clusters as a function of cluster age for clusters from the LMC.
Since each cluster is an SSP, this plot effectively shows the $V-K$
evolution of a single SSP at the mean metallicity of the LMC, which is
$Z=0.006$ \citep{Cioni06}.  This figure includes the SPS predictions
for a standard model at $Z=0.006$, and models where the parameters
$\Delta_T$ and $\Delta_L$ and varied by $\pm 0.2$ and $\pm 0.4$,
respectively. 

As noted in \citet{Marigo08}, the SSP rapidly reddens at $t\approx
10^8$ years because the AGB phase occurs only in stars with masses
less than $M_i\sim 5\, \Msun$.  \citet{Marigo08} also note that this
rapid reddening is stronger than observations allow, and they suggest
that the source of the discrepancy lies in the treatment of mass-loss
in the AGB phase.  We mention in passing that the models of
\citet{Maraston05} do not suffer from this effect
\citep[see][]{Maraston98}, in part because they were calibrated to fit
the LMC star cluster data.  At ages younger than $10^8$ years the
model predicts colors too blue compared to observations.  Such young
clusters are not thought to contain AGB stars; the discrepancy can
thus not be attributed to the AGB phase.  \citet{Marigo08} offer a
number of possible explanations for the mismatch between model and
data but ultimately this mismatch must be accepted in the current
generation of models.

Moreover, it is clear that the default models (where $\Delta_L=0.0$
and $\Delta_{T}=0.0$) predict $V-K$ colors that are too red compared
to the observed star clusters at $t\gtrsim10^8$ years.  The figure
shows that increasing the temperature of TP-AGB stars by 0.2 dex
and/or decreasing their luminosity by 0.4 dex results in better
agreement with the data.  Models that decrease the temperature and/or
increase the luminosity do not produce substantially worse agreement
than the default model at late times ($t\gtrsim10^{9.5}$ years) but
are much redder than the data for intermediate ages.  

From this comparison we conclude that values for the TP-AGB parameters
in the range $-0.4<\Delta_L<0.0$ and $0.0<\Delta_{T}<0.2$ cannot be
ruled out by current data from the LMC.  When fitting to observations
we allow these parameters to vary in the wider range
$-0.4<\Delta_L<0.4$ and $-0.2<\Delta_{T}<0.2$ for increased
flexibility because these parameters may depend on metallicity in a
way that could not be discerned from the comparison to the LMC. 

\begin{figure}[t!]
\plotone{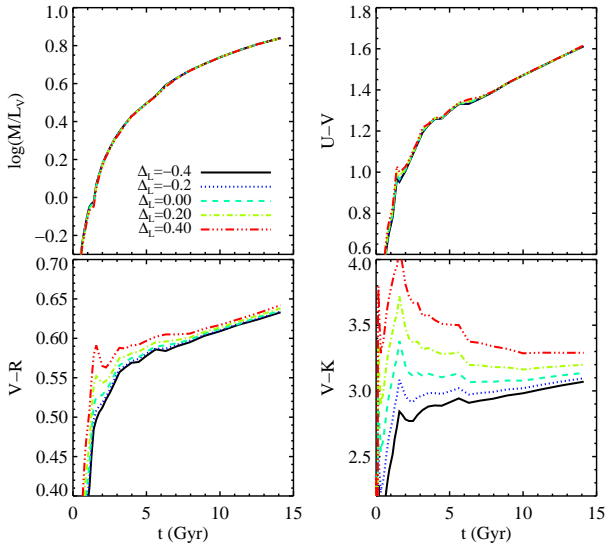}
\vspace{0.5cm}
\caption{Evolution of SSPs at solar metallicity for a range in the
  parameter $\Delta_L$, the overall shift in ${\rm log}(L_{\rm bol})$
  of the TP-AGB phase.  Since TP-AGB stars are cool, varying
  $\Delta_L$ primarily effects bands redward of $R$.  The
  mass-to-light ratio in the $V$-band, $M/L_V$, and the $U-V$ color
  are entirely insensitive to this parameter.}
\label{fig:dell}
\vspace{0.5cm}
\end{figure}

\begin{figure}[t!]
\plotone{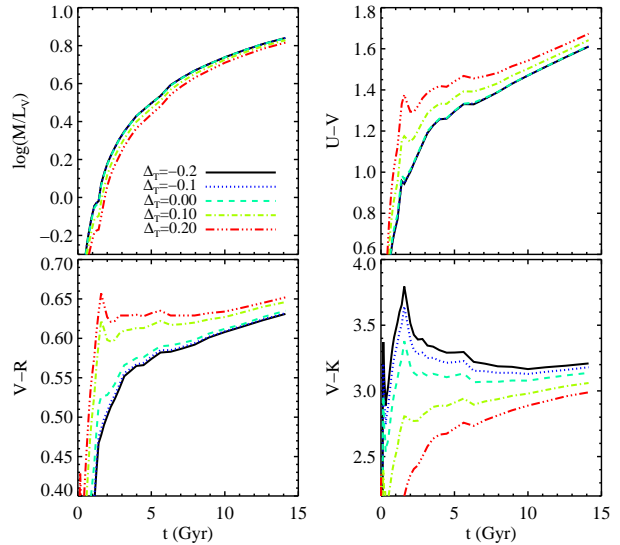}
\vspace{0.5cm}
\caption{Same as Figure \ref{fig:dell}, except now the parameter
  $\Delta_T$, the overall shift in ${\rm log}(T_{\rm eff})$ of the
  TP-AGB phase, is varied.  Increasing the temperature of the TP-AGB
  phase effectively shifts the dominant flux contribution of these
  stars from the $K$-band to bluer bands such as $R$ and $V$.  Thus
  hotter TP-AGBs can result in redder $U-V$ colors as shown in the
  figure.}
\label{fig:delt}
\vspace{0.5cm}
\end{figure}

The effects of $\Delta_L$ and $\Delta_{T}$ on the evolution
of SSPs are explored more broadly in Figures \ref{fig:dell} and
\ref{fig:delt}, where the evolution of $M/L_V$, $U-V$, $V-R$, and $V-K$
are plotted for a range of these TP-AGB parameters.

Figure \ref{fig:dell} shows that it is primarily bands redward of $V$
that are sensitive to $\Delta_L$ --- bluer bands are almost entirely
insensitive to this parameter.  This result is not surprising in light
of the fact that TP-AGB stars dominate the energy output redward of
$\sim1\mu m$.  It thus may be the case that the inclusion of near-IR
information does not significantly reduce the errors in physical
properties of galaxies, such as stellar masses, since the red light in
galaxies is dominated by this uncertain phase of stellar evolution.

Figure \ref{fig:delt} shows the impact of the $\Delta_{T}$
parameter on the evolution of SSPs.  This TP-AGB parameter has a
strong influence on not only the $V-K$ color but on bluer colors as
well.  Moreover, this parameter has opposite effects on $V-K$ compared
to $U-V$ and $V-R$ colors, in the sense that $\Delta_{T}>0$
results in bluer $V-K$ colors and redder $U-V$ and $V-K$ colors.
This trend is due to the fact that hotter TP-AGB stars emit most of
their light in progressively bluer bands compared to cooler TP-AGB
stars.  So as $\Delta_{T}$ is increased, the integrated flux
in the $V-$ and $R-$bands is increased at the expense of redder bands
such as $K$.

These TP-AGB parameters will be included in our fits to observed
broad-band photometry of galaxies in $\S$\ref{s:res}.

\subsection{Horizontal branch stars}
\label{s:hb}

The horizontal branch (HB) consists of old low-mass
($M\lesssim1\Msun$) stars burning helium in their cores, and hydrogen
in a shell surrounding the core, at nearly constant bolometric
luminosity \citep{Sweigart87, Lee90, Lee94}.  The narrow range of
luminosities is driven largely by the constant core He mass of stars
that enter the HB phase.

The temperature of HB stars has received extensive attention in the
literature, in part because HB stars are luminous and thus accurate
knowledge of their temperatures is important for modeling the
integrated light from a galaxy.  Observationally, it appears that
metal-rich globular clusters  contain almost exclusively red HB
stars, while HB stars in metal-poor clusters ([Fe/H]$\lesssim-1.4$)
populate a wide range of temperatures \citep{Harris96}.  This trend is
thought to be due to metallicity-dependent mass-loss on the RGB, which
leads to a smaller envelope mass in metal-poor compared to metal-rich
systems, and hence hotter HB stars \citep{Greggio90}.  One
of the fundamental problems with understanding the morphology
(i.e. temperature distribution) of the HB is that the temperature of
HB stars is extremely sensitive to the amount of mass lost.  For
example, a difference in envelope mass of only $0.04\Msun$ can
correspond to a temperature increase from $5,000$ K to $11,000$ K
\citep{Rich97}, a temperature increase sufficient to explain the
morphology of metal-poor systems.  Since an adequate theory of
mass-loss for evolved stars does not exist, understanding the
morphology of the HB from first principles is extremely challenging.

One might hope to understand the HB morphology purely observationally
and thereby asses its potential impact on SPS without the aid of
theoretical models.  This approach has also proved difficult.  In
particular, there are clear examples of metal-rich star clusters with
a blue HB \citep{Rich97, Brown00, Kalirai07}, suggesting that the
morphology of the HB may not depend only on metallicity.  On the other
hand, data from the {\it Hipparcos} satellite demonstrate that there
is not a significant population of blue HB stars in the solar
neighborhood \citep{Jimenez98}.  This may to attributed to the paucity
of metal-poor stars in the solar neighborhood \citep{Rana91}.
However, we might expect other galaxies, especially ellipticals, to
contain a significant population of metal-poor stars because recent
galaxy formation models imply that ellipticals form their stars in a
manner closely analogous to the closed-box model.  Indeed SPS modeling
of the spectra of local elliptical galaxies suggests that $\sim5-20$\%
of the stellar mass in such galaxies may be associated with metal-poor
stars \citep{Worthey96, Maraston00, Trager05, Maraston09}.

\begin{figure}[t!]
\plotone{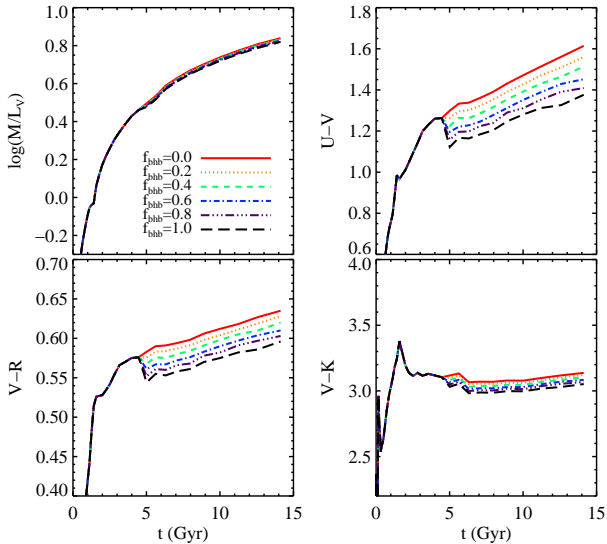}
\vspace{0.5cm}
\caption{Same as Figure \ref{fig:dell}, except now the parameter
  $f_{BHB}$, the fraction of blue HB stars, is varied.  In our model
  this parameter is turned on at $t=5$ Gyr, as is clear from the
  figure, because there are no observed globular clusters with a blue
  horizontal branch and an age of less than 5 Gyr.  Blue HB stars are
  hot; increasing their fraction thus has the largest impact on bluer
  colors such as $U-V$ and $V-R$, and a smaller effect on $V-K$.}
\label{fig:bhb}
\vspace{0.5cm}
\end{figure}

Blue HB stars are an important phase of stellar evolution to
understand because they can strongly effect the balmer lines
\citep[e.g.][]{LeeHC00}, and can complicate the estimation of
quantities such as stellar age and metallicity
\citep[e.g.][]{LeeHC02}.  With sufficient spectral resolution and
signal-to-noise, the spectral signatures of a blue HB can in fact be
isolated \citep{Trager05}.  However, for broad-band colors and low
signal-to-noise data, it is difficult to make this distinction.

Despite these large observational and theoretical uncertainties, some
SPS codes do not include blue HB stars \citep{Bruzual03}, or, if they
are included, they are included only in metal-poor systems
\citep{Jimenez04}.  Only the \citet{Maraston05} model includes blue HB
stars at both low and high metallicity.  We feel that the
uncertainties in our knowledge of the HB warrant a more flexible
treatment of the HB than is typically implemented.

To this end, in our SPS model we allow for a specified fraction of HB
stars that are spread uniformly in temperature from the standard red
end of the HB to $10,000$ K {\it for the full range of metallicities
  we consider}.  We refer to these HB stars with temperatures hotter
than the red clump as blue HB stars.  These blue HB stars are added
for populations older than 5 Gyr.  No globular clusters have been
found with both younger ages and a blue HB morphology.
\citet{Atlee09} have recently shown that the UV excess in luminous
early-type galaxies does not evolve appreciably since $z\approx0.65$,
suggesting that, if the UV excess is due to blue HB stars, then it is
reasonable to assume a non-evolving blue HB for stellar populations
older than several gigayears.

\begin{figure}[t!]
\plotone{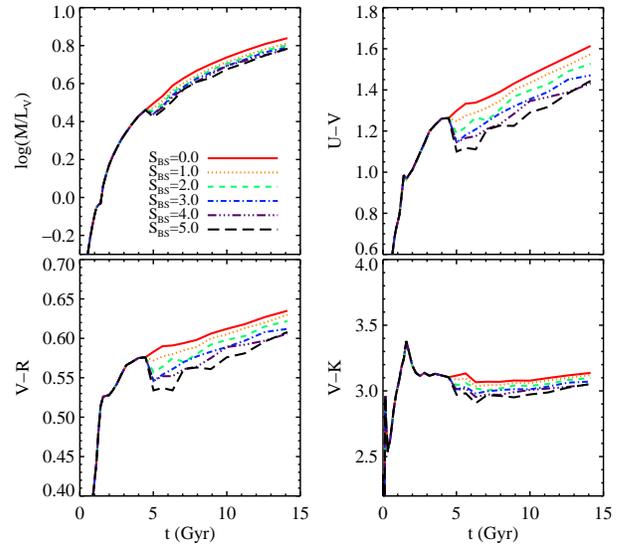}
\vspace{0.5cm}
\caption{Same as Figure \ref{fig:dell}, except now the parameter
  $S_{BS}$, the specific frequency of blue stragglers, is varied.  In
  our model this parameter is turned on at $t=5$ Gyr, as is clear from
  the figure.  The impact of BS stars is qualitatively similar to but
  smaller in magnitude than blue HB stars because BS stars are at
  roughly the same temperature as blue HB stars but are nearly an
  order of magnitude less luminous (see Figure \ref{fig:obshr}).}
\label{fig:sbs}
\vspace{0.5cm}
\end{figure}

The bolometric luminosity of HB stars is left unaltered because, as
stated above, their luminosities are understood both observationally
and theoretically.  The weight assigned to blue HB stars is a free
parameter and is specified as the fraction of HB stars that are in
this blue component, $f_{BHB}$.  This parameter is allowed to vary
from $0<f_{BHB}<0.5$.  The upper limit is arbitrary but is motivated
by the consideration that a galaxy with $20$\% of its stellar mass in
metal-poor stars and a blue HB population in $40$\% of its metal-rich
systems would have $f_{BHB}=0.5$.  Observationally, \citet{Dorman95}
infer from the UV upturn in elliptical galaxies that between $5-20$\%
of HB stars are very blue.  One might expect this fraction to increase
as the overall metallicity of a galaxy decreases.

Figure \ref{fig:bhb} shows the color and $M/L$ evolution of SSPs at
solar metallicity for a range in the parameter $f_{BHB}$.  The bluer
colors are more sensitive to this parameter than the redder colors.
The addition of blue HB stars, in our implementation, does not
significantly alter the evolution of colors at late times, but rather
it amounts to an approximately constant blueward offset.

\begin{figure}[t!]
\plotone{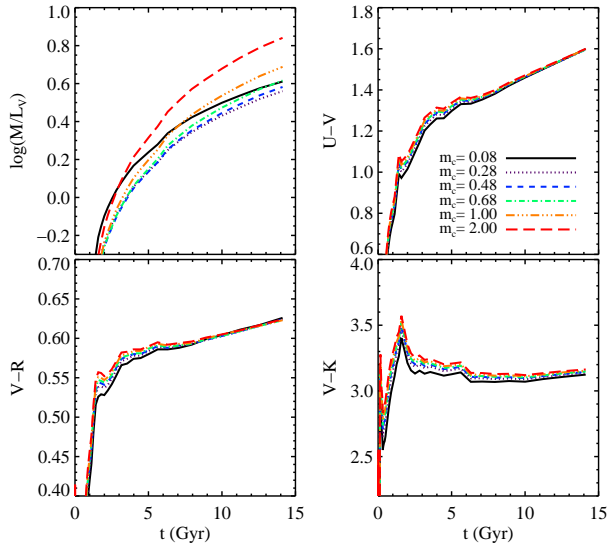}
\vspace{0.5cm}
\caption{Same as Figure \ref{fig:dell}, except now the parameter
  $m_c$, the characteristic mass of the IMF, is varied.  As discussed
  in the text, the IMF has a much larger influence on the luminosity
  evolution, and hence $M/L$, than color evolution.  Nonetheless, IMF
  differences can induce color changes of order $0.1$ mags in the
  near-IR at intermediate ages.}
\label{fig:vdmc}
\vspace{0.5cm}
\end{figure}

\subsection{Blue straggler stars}

As indicated in Figure \ref{fig:obshr}, blue stragglers (BSs) are
stars that extend blueward and brighter than the main sequence
turn-off point.  They are a ubiquitous feature of observed globular
clusters \citep[e.g.][]{Sandage53, Bailyn95, Brown05, Sarajedini07}.
Their origin remains a mystery.  Standard explanations are that BSs
are either due to primordial binary evolution \citep{McCrea64} or
collisional merging \citep{Bailyn95}.  If the later is in fact
dominant then one expects BSs to have a negligible effect on the
integrated light from a galaxy because collisions are only important
at the high densities found in globular clusters, in which only a
small fraction of galactic stellar mass is found.  If binary
evolution, due e.g. to mass transfer and/or coalescence, is
responsible, then they may be common throughout galaxies due to the
high primordial binary fraction of field stars of spectral type $\sim
K$ or earlier \citep{Duquennoy91, Lada06}.  Observations seem to
suggest that both processes are at work in globular clusters
\citep{Davies04, Piotto04}.

There are also some observational indications that BSs are factors of
several more common in the field compared to globular clusters
\citep{Preston00}.  There are several ways of explaining such a
difference between field and cluster, but it is sufficient for our
purposes to note that their observational status is sufficiently
underconstrained to include BSs in our SPS modeling in a flexible way.

Turning to their observational consequences, \citet{Li09a} have
incorporated the effects of binary star interactions into their SPS
code and find that including binaries can result in broad-band colors
bluer by $\sim0.05$ mags.  \citet{Xin05} find that BSs in old open
clusters can contribute as much as $\sim0.2$ magnitudes to the $B-V$
color of the integrated cluster.  They find no correlation between the
number of BSs and the cluster age, suggesting that BSs may be a
relatively stable phenomenon \citep[though see discussion
in][]{Davies04}.

BSs are included in our SPS code with the following prescription.  As
with the blue HB stars, the BS phenomenon is `turned on' for ages
greater than 5 Gyr.  Observationally, the specific frequency of
BSs, $S_{BS}$, is defined as the number of BSs per unit HB star,
i.e. $S_{BS}\equiv N_{BS}/N_{HB}$.  Typical values for globular
clusters are in the range $0.1<S_{SB}<1$ \citep{Piotto04}, although
\citet{Preston00} suggest that $S_{BS}$ could be as high as five in
the field.  BSs typically are found in a region parallel to the
zero-age main sequence starting at $\sim0.5$ mags brighter than the
turn-off point and extending another $\sim2$ mags brighter
\citep{Xin05}.  In our code BSs populate a region from $0.5-2.5$ mags
brighter than the zero-age main sequence.  The BS population is
parameterized with the single free parameter $S_{BS}$.

Figure \ref{fig:sbs} shows the effect of BSs on the integrated colors
and $M/L$ of a solar metallicity SSP.  BSs contribute mostly to the
blue bands and at the $\sim0.1$ mag level, consistent with both models
\citep{Li09a} and observations \citep{Xin05}.  BSs have less of an
effect than HB stars because they are roughly an order of magnitude
less luminous (cf. Figure \ref{fig:obshr}).

\subsection{The IMF}
\label{s:uimf}

\begin{figure}[t!]
\plotone{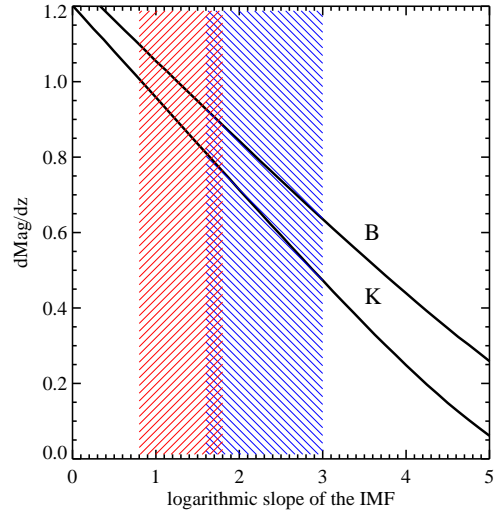}
\vspace{0.5cm}
\caption{Change in magnitude per unit redshift at $z\leq1$ for a
  passive stellar population, as a function of the logarithmic slope
  of the IMF.  Magnitude evolution in both the $B-$ and
  $K-$bands are included.  The hatched regions denote the
  uncertainties in the measured IMF slope at $0.08<M<0.5\Msun$ ({\it
    left}) and $M\ge1.0\Msun$ \citep[{\it right};][]{Kroupa01}.  Since
  luminosity evolution is determined by the IMF slope at the mass
  where stars are just leaving the main sequence, passive evolution is
  determined by the slope at $\approx 1\Msun$ for old
  ($t\gtrsim10^{10}$ yrs) stellar populations.  In the $K-$band, the
  uncertainty in the IMF slope at $1\Msun$ translates into an
  uncertainty in $dM/dz$ of $\sim0.4$ mags for an old, passively
  evolving stellar population.}
\label{fig:lumev}
\vspace{0.5cm}
\end{figure}

The IMF is very uncertain at masses $0.8\lesssim M\lesssim2 \Msun$
because at these masses IMF constraints derived from field stars in
the solar neighborhood rely on large and uncertain stellar evolution
corrections \citep{Kroupa01}.  While the IMF is also uncertain at both
higher and lower stellar masses, this mass range is critical for
galaxies made of old ($\gtrsim 10^9$ years) stars because the
main-sequence turn-off of such old populations is
$\sim1\Msun$\footnote{The main-sequence turn-off mass decreases very
  slowly with time for old stellar populations.  For example, at
  $t=10^9, 5\times10^9$, and $10^{10}$ years the turn-off mass is
  $2.0, 1.2$, and $1.0\Msun$, respectively.}, and, as discussed in
$\S$\ref{s:imf}, it is the logarithmic slope of the IMF at the
main-sequence turn-off point that determines the rate of luminosity
evolution for a passively evolving system.  More generally, the
observed IMF must be corrected for a large number of systematics
including the fraction of binary stars, shot noise, and dynamical
evolution, in order to infer the `true' underlying single-star IMF.
These corrections are large and uncertain.

While direct measurements of the IMF in the Galaxy are challenging,
they are all but impossible in external galaxies beyond the LMC and
SMC.  This unfortunate situation is unlikely to improve.  Nonetheless,
much work has gone into constraining the IMF via indirect approaches.
For example, evolution with redshift of $M/L_B$ and $U-V$ for galaxies
on the fundamental plane suggests that the IMF may be weighted toward
higher masses at $z\sim2$ compared to local measurements
\citep{vanDokkum08}.  Furthermore, comparison of the evolution of the
cosmic star formation rate density, which is sensitive to the
high-mass end of the IMF, to evolution of the cosmic stellar mass
density, which is sensitive to the lower-mass end, also suggests that
the IMF is different at $z>1$, in the same sense as implied by the
evolution of the fundamental plane \citep[e.g.][]{Hopkins06b, Dave08,
  Wilkins08}.  Carbon-enhanced metal poor-stars also point toward a
top-heavy IMF at early times \citep{Lucatello05, Tumlinson07a,
  Tumlinson07b}.  These arguments are indirect, and should thus be
treated with caution.  However, in support of these conjectures,
simple theoretical arguments suggest that the characteristic
temperature of molecular clouds may be higher at earlier epochs, which
in turn would lead to a higher Jeans mass and thus potentially a
different IMF at early times \citep{Larson98, Larson05}.  For a review
of evidence suggesting a varying IMF, see \citet{Elmegreen09}.

In standard implementations of SPS the IMF is typically treated as an
uninteresting overall normalization of the galactic stellar mass.
This assumption is problematic for at least two related reasons as
demonstrated in Figure \ref{fig:vdmc}, and first pointed out by
\citet{Tinsley80}.  On the one hand, the IMF determines the relative
weights to be assigned to various portions of the isochrones.  If
broad-band colors are dominated by one phase of stellar evolution, for
example the main sequence turn-off, then they will be entirely
insensitive to the IMF.  This is approximately the case for the $U-V$
color in Figure \ref{fig:vdmc}.  However, if at least some colors are
determined by a mixture of stellar evolutionary phases, such as the
RGB and AGB phases, then these colors will be sensitive to the IMF, as
seen in the $V-K$ color in the figure \citep[see also][]{Maraston98}.

This uncertainty in the colors due to the uncertainty in the IMF
should arguably be included in SPS models because observational
quantities are effected.  Different IMFs can yield different colors
which in turn will result in different predictions for ages and
metallicities of galaxies.  This can be contrasted with the effect of
the IMF on the stellar mass of galaxies because there variation in the
IMF leads only to an overall re-scaling of the stellar mass.  See
\citet{Westera07} for a discussion of these effects and the prospects
for constraining the form of the IMF from broad-band photometry of
external galaxies.  In the present work we do not marginalize over
uncertainties in the IMF when fitting the photometry of observed
galaxies.

The second issue is that, even if colors are only mildly affected, the
rate of luminosity evolution is strongly related to the logarithmic
slope of the IMF, as shown in the top left panel of Figure
\ref{fig:vdmc}.  So while the stellar mass of galaxies {\it at a fixed
  epoch} may be relatively robust against IMF variations, in the sense
that changing the IMF will change the stellar mass of all galaxies by
the same relative amount, luminosity evolution is very sensitive to
the IMF.  The slope of the IMF near $1 \Msun$ is not known to better
than $\pm (0.3-0.7)$ {\it in the solar neighborhood} \citep{Kroupa01},
and thus this is an important uncertainty to consider even if one
wants to retain the notion of a universal IMF.  This uncertainty makes
it extremely difficult, for example, to interpret evolution in the
luminosity function of bright galaxies in terms of a constraint on
their stellar growth.

This point is demonstrated explicitly in Figure \ref{fig:lumev}, where
we plot the best-fit linear slope for the evolution of the absolute
magnitude of an SSP with redshift, over the interval $0<z<1$, as a
function of the logarithmic IMF slope.  As mentioned above, for old
stellar populations the main sequence turn-off mass is $\sim1\Msun$,
and it is the slope of the IMF at the turn-off mass that is important
for passive luminosity evolution.  The figure demonstrates that
passive fading of a population of stars since $z=1$ is uncertain at
the $\sim0.4$ mag level in the $K$-band \citep[see also][]{Yi03}.  In
light of this effect, precise conclusions made from evolving
luminosity functions \citep[e.g.][]{Wake06, Brown07, Cool08} must be
treated with extreme caution.  See $\S$\ref{s:whyimf} for a detailed
discussion of this issue.

\subsection{Metallicity distributions}

\begin{figure}[t!]
\plotone{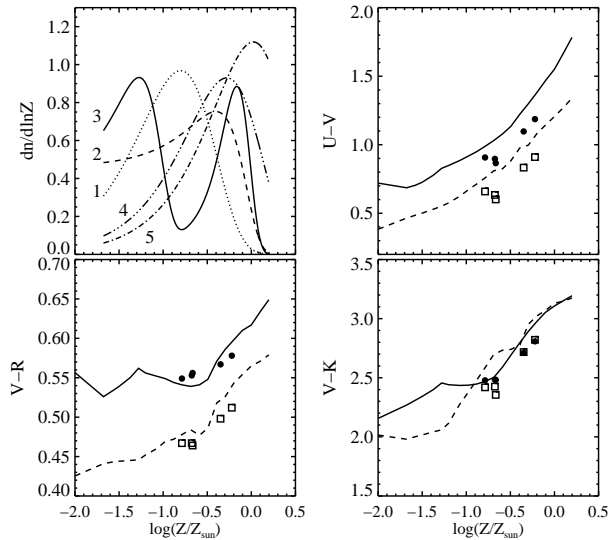}
\vspace{0.5cm}
\caption{Effect of stellar metallicity on the evolution of an SSP.
  This figure compares the effect of a {\it distribution} of
  metallicities on the colors of SSPs to the colors of
  single-metallicity SSPs.  Five metallicity distributions are shown
  in the top left panel.  The colors of SSPs with such metallicity
  distributions are shown in the remaining three panels at $t=3.2$
  ({\it open squares}) and $12.6$ Gyr ({\it filled circles}), as
  a function of the mean metallicity of the distribution.  These
  results are compared to the colors of single-metallicity SSPs as a
  function of metallicity, at $3.2$ ({\it dashed lines}) and $t=12.6$
  Gyr ({\it solid lines}).  The distributions in the upper left panel
  are numbered according to their mean metallicities, in ascending
  order.  Notice that the difference between the points and lines is
  small for colors redward of the $V-$band but are noticeable and
  systematic for $U-V$.  See the text for details.}
\label{fig:vmetals}
\vspace{0.5cm}
\end{figure}

The stars within a galaxy do not form from gas at a single
metallicity.  Rather, if stars form from a ``closed-box'' of gas,
where no inflow or outflow takes place, the metallicity distribution
of the stars approaches:
\noindent
\be
\frac{{\rm d}n}{{\rm dln}Z} \propto Z\,e^{-Z/p}
\label{e:cbox}
\ee
\noindent
where $p$ is the true yield \citep{Searle72, Pagel97}.  It has been
known for some time that the closed-box model predicts too many
metal-poor stars in the solar neighborhood --- the so-called G-dwarf
problem \citep{Rana91}.  A possible explanation is that gaseous inflow
and/or outflow has been important in the formation of stars in the
solar neighborhood \citep{Larson72}.  Such an explanation is
particularly attractive in the context of hierarchical structure
growth, which is a robust feature of a universe dominated by cold dark
matter.

In spite of the failure of the closed-box model at describing the
solar neighborhood, one might expect it to fair better at describing
the metallicity distribution of stellar populations formed over short
time-scales.  In such populations gaseous inflow and/or outflow are
less important because they have less time to effect the metal content
of the gas out of which stars are forming.  Indeed, the metallicity
distribution of stars in the bulge of the Milky Way can be somewhat
better-described by a closed-box model with a yield of $p=0.015$,
although even in the bulge there appears to be too few metal-poor
stars \citep{Rich90, Zoccali03}.  Inclusion of the metal-poor halo in
the accounting may yet produce better agreement with the closed-box
model.  The paucity of metal--poor stars in comparison to the
predictions of the closed--box model is evident in other galaxies as
well \citep{Worthey96, Worthey05}.

While it is clear that there are fewer observed metal-poor stars
compared to the closed-box model, the stars constituting galaxies
nevertheless have a distribution of metallicities and this fact is not
included in standard SPS modeling.  In order to explore the effects of
a distribution of metallicities on the colors of SSPs, we consider
three closed box models described by Equation \ref{e:cbox} with
$p=0.003, 0.010, 0.020$, and two more flexible distributions that are
the sum of two Gaussians:
\noindent
\be
\frac{{\rm d}n}{{\rm dln}Z} \propto e^{-(Z-\mu_1)^2/\sigma_1^2/2} + e^{-(Z-\mu_2)/\sigma_2^2/2}
\ee
\noindent
where $(\mu_1,\sigma_1^2,\mu_2,\sigma_2^2)=(0.010,0.005,0.005,0.005)$
and $(0.013,0.003,0.0005,0.001)$.  In light of the discussion above,
these distributions almost certainly contain more metal-poor stars
than observations demand, and so they should be thought of as extreme
scenarios for the metallicity distributions in real galaxies.  Note
also that we are assuming that metallicity and stellar age are
uncorrelated.  Of course, the mean metallicity of a galaxy will tend
to increase with time and so one might expect younger stars to be more
metal rich, on average, than older stars.  Investigating constraints
on the metallicity evolution of a galaxy will be the subject of future
work.

The five metallicity distributions are shown in the top left panel of
Figure \ref{fig:vmetals}.  In the remaining panels of this figure we
plot the dependence of the $U-V$, $V-R$, and $V-K$ colors on
metallicity at two stellar ages.  The colors as a function of
single-metallicity SSPs ({\it lines}) are compared to the colors from
the multi-metallicity stellar populations described above ({\it
  symbols}).  Colors from the multi-metallicity populations are
plotted against the average metallicity of the population.  For the
$V-R$ and $V-K$ colors there is no significant difference between a
single-metallicity SSP and a multi-metallicity SSP with the same
average metallicity.  Since we have chosen rather extreme metallicity
distributions, this implies that the optical and near-IR colors shown
here are entirely insensitive to the range of metallicities in a galaxy.

For the $U-V$ color the multi-metallicity populations are $\sim0.1$
magnitudes bluer when compared to single-metallicity SSPs.  The impact
of a distribution of metallicities on the inferred color is
comparatively strong for $U-V$ because this color varies strongly, and
non-linearly, with metallicity.  We have found that this discrepancy
between single and multi-metallicity populations increases into the
ultraviolet \citep[see also][]{Schiavon07}.  We leave a more detailed
discussion of this effect for future work, but note here that this
difference could potentially lead to a mis-estimation of either/both
the metallicity and age of a galaxy.

Recall that we have decoupled metallicity and horizontal branch
morphology.  As noted in $\S$\ref{s:hb}, metal-poor globular clusters
have extended horizontal branches, while metal-rich clusters have
exclusively red horizontal branch stars.  This can certainly produce a
substantial difference between a single metallicity and
multi-metallicity population of stars.  Since we consider the
horizontal branch morphology separately herein, we do not include its
effect in this section.  We have decoupled metallicity from horizontal
branch morphology because, as discussed in $\S$\ref{s:hb}, the
dependence of the horizontal branch on metallicity is poorly
understood.

\subsection{Summary of important unknowns}

In this section we have explored several important and yet uncertain
ingredients in SPS modeling.  These include uncertain phases of
stellar evolution such as the TP-AGB stars, horizontal branch, and
blue stragglers, the IMF, and the universal assumption in SPS modeling
that every star in a galaxy has the same metallicity.

As Figures \ref{fig:dell}, \ref{fig:delt}, \ref{fig:bhb}, and
\ref{fig:sbs} show, the uncertain phases of stellar evolution can have
a large effect on the color of an SSP within the plausible range of
uncertainties.  These uncertainties thus must be incorporated into any
SPS model in order to derive accurate and robust physical parameters
of galaxies.

Figures \ref{fig:vdmc} and \ref{fig:lumev} show that the variations in
the logarithmic slope of the IMF has a minor effect on the colors of
an SSP but has a large effect on its luminosity evolution.  This
uncertainty must thus be incorporated whenever one attempts to discuss
the luminosity evolution of galaxies.

Finally, Figure \ref{fig:vmetals} compares the effect of assuming a
single metallicity to a distribution of metallicities on the evolution
of an SSP.  The relation between color and average metallicity is
essentially independent of the distribution of metallicities one
assumes for bands redward of $V$.  At bluer wavelengths the difference
between single and multi-metallicity populations becomes more
apparent.  This effect, if not included in SPS models, may introduce
unknown systematics.  We will investigate this aspect in future work.

\begin{figure}[t!]
\plotone{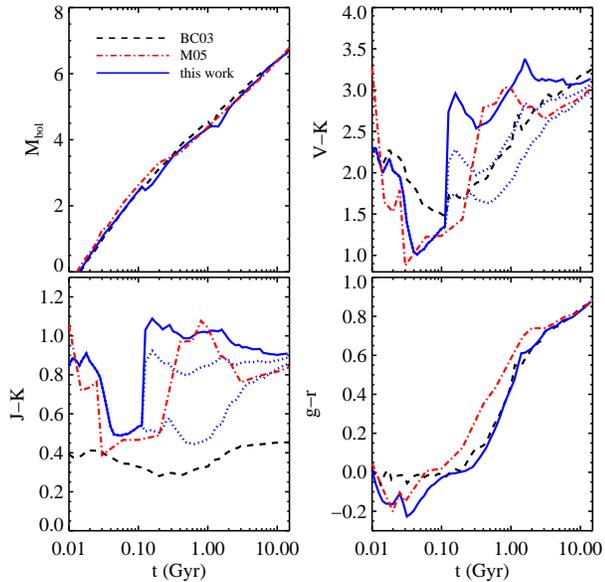}
\vspace{0.5cm}
\caption{Comparison of the \citet[][B03]{Bruzual03} and
  \citet[][M05]{Maraston05} SPS models to the default model presented
  herein for a solar metallicity SSP.  We also include results of our
  model with $\Delta_L=-0.4$ ({\it upper dotted line}) and
  $\Delta_T=0.2$ ({\it lower dotted line}) in the panels where these
  parameter choices result in noticeably different model predictions.}
\label{fig:compare}
\vspace{0.5cm}
\end{figure}

\subsection{Comparison to other SPS models}

In this section we compare our SPS model to two previous, highly used,
models.  Figure \ref{fig:compare} shows the luminosity and color
evolution of a solar metallicity SSP for our default model and the
models of \citet[][BC03]{Bruzual03} and \citet[][M05]{Maraston05}.
Also included in the figure are two predictions of our model with the
TP-AGB parameters set to $\Delta_L=-0.4$ and $\Delta_T=0.2$.  The M05
model predicts that the TP-AGB stars constitute a much larger fraction
of the near-IR light compared to BC03.  This explains the much redder
$J-K$ and $V-K$ colors, and the redder $g-r$ colors for $t\lesssim 2$
Gyr shown in Figure \ref{fig:compare}.

Our default SPS model compares favorably with the M05 model, which is
not surprising because our isochrones include TP-AGB calculations of a
sophistication comparable to that in M05.  In addition, our code is
flexible enough to roughly capture the trends predicted by BC03.  In
the figure it is clear that the $V-K$ color evolution of our model
appears similar to BC03 when we set $\Delta_L=-0.4$.  In other words,
when we force the TP-AGB stars to be less important to the integrated
SSP, our model converges to the BC03 model.  This is one of the
advantages of the flexible SPS model presented herein compared to
others in the literature --- the flexible model does not put priors on
uncertain physics in stellar evolution.  Instead, the model lets the
data decide what parameters are most favored.

\begin{deluxetable}{lll}
\tablecaption{Definition of SDSS samples}
\tablehead{ \colhead{Description} & \colhead{Luminosity Cut} &
\colhead{Color Cut} }
\startdata\\
bright red & $-24.0<M_r-5{\rm log}(h)<-22.0$ & $0.70<(g-r)<0.90$   \\
faint red & $-20.5<M_r-5{\rm log}(h)<-20.0$ & $0.70<(g-r)<0.90$   \\
bright blue & $-21.5<M_r-5{\rm log}(h)<-21.0$ & $0.20<(g-r)<0.66$   \\
faint blue & $-20.5<M_r-5{\rm log}(h)<-20.0$ & $0.20<(g-r)<0.66$   \\
\enddata
\label{t:sdss}
\tablecomments{All samples are within the redshift range
  $0.10<z<0.12$.}
\end{deluxetable}

\section{Data, and Data Fitting}
\label{s:data}

We now turn to the task of fitting our SPS model to the photometry of
galaxies.  In this section we describe the data used and the details
of the fitting procedure.  In the following section we present the
results.

\subsection{Data at $z\sim0$}

The Sloan Digital Sky Survey \citep[SDSS;][]{York00,Abazajian04,DR4}
is an extensive photometric and spectroscopic survey of the local
Universe.  As of Data Release 4 (DR4), imaging data exist over $6670$
deg$^2$ in five band-passes, $u$, $g$, $r$, $i$, and $z$.
Approximately $670,000$ objects over $4780$ deg$^2$ have have been
targeted for follow-up spectroscopy as part of SDSS are included in
DR4; most spectroscopic targets are brighter than $r=17.77$
\citep{Strauss02}.  Automated software performs all the necessary data
reduction, including the assignment of redshifts.

The Two Micron All Sky Survey \citep[2MASS;][]{Jarrett00, Jarrett03}
is an all-sky map in the $J$, $H$, and $K_s$ bands.  We make use of
the Extended Source Catalog, which is complete to $K_s\sim13.5$.  The
2MASS and SDSS catalogs are matched as described in \citet{Blanton05},
and are available in the hybrid NYU Value Added Galaxy Catalog
\citep[VAGC;][]{Blanton05}.

In addition we use the publicly available package \texttt{kcorrect
  v4.1.4} \citep{Blanton03b, Blanton07} to derive restframe $ugrizJHK$
Petrosian magnitudes for all SDSS galaxies that have a matched 2MASS
counterpart.  All galaxies are K-corrected to $z=0.0$.  The
\texttt{kcorrect} package also estimates stellar masses by fitting the
broad-band colors to a grid of templates based on the
\citet{Bruzual03} SPS code; see \citet{Blanton07} for details.  In
$\S$\ref{s:res} we will compare to these mass estimates.  In the
present work we are interested in deriving physical parameters for a
small but representative sample of galaxies, and so we are not
concerned with issues such as survey completeness or the construction
of volume limited samples.

Four samples from the SDSS-2MASS matched catalog are defined in order
to explore trends with luminosity and color.  All galaxies are chosen
to have $0.10<z<0.12$ in order to minimize any possible evolutionary
effects within the sample.  The samples are defined according to the
luminosity and color cuts defined in Table \ref{t:sdss}.

\begin{figure*}[t!]
\plotone{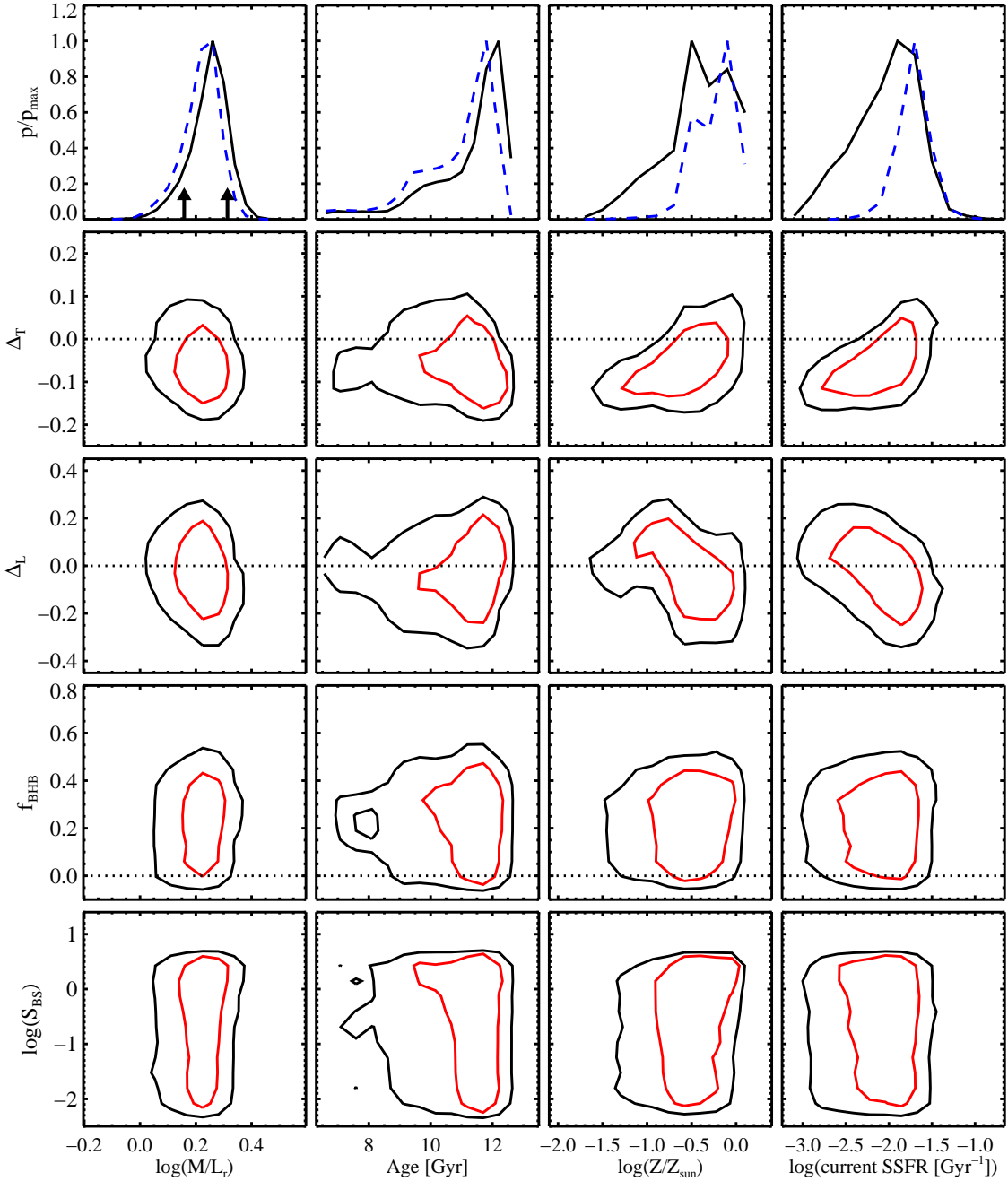}
\vspace{0.5cm}
\caption{Results from fitting our SPS model to the broad-band
  photometry of an $\sim L^\ast$ blue galaxy at $z\sim0$. {\it Top
    Row:} Marginalized probability distributions for physical
  properties including $M/L$, mass-weighted age, metallicity, and
  current specific SFR.  Results are included both for our default
  model with no marginalization over uncertain aspects of stellar
  evolution ({\it dashed lines}) and a model that explicitly
  marginalizes over such uncertain aspects ({\it solid lines}).  The
  arrows in the left-most panel indicate the best fit $M/L$ for this
  galaxy as determined by different authors (see the text for
  details).  {\it Bottom Rows:} Likelihood contours at 68\% CL ({\it
    red lines}) and 95\% CL ({\it black lines}) highlighting the
  dependence of the four physical properties on four parameters
  governing uncertain aspects of stellar evolution (see Table
  \ref{t:models} for details).  In most cases the upper and lower
  bounds on the parameters along the $y$-axis are governed by the
  choice of priors.  The dotted lines highlight the values of the
  parameters assumed in the default model.  The default value for the
  quantity along the bottom row, ${\rm log}(S_{BS})$, is $S_{BS}=0$
  and hence is not visible in the figure.}
\label{fig:cgrid_bblu}
\vspace{0.5cm}
\end{figure*}

\begin{figure*}[t!]
\plotone{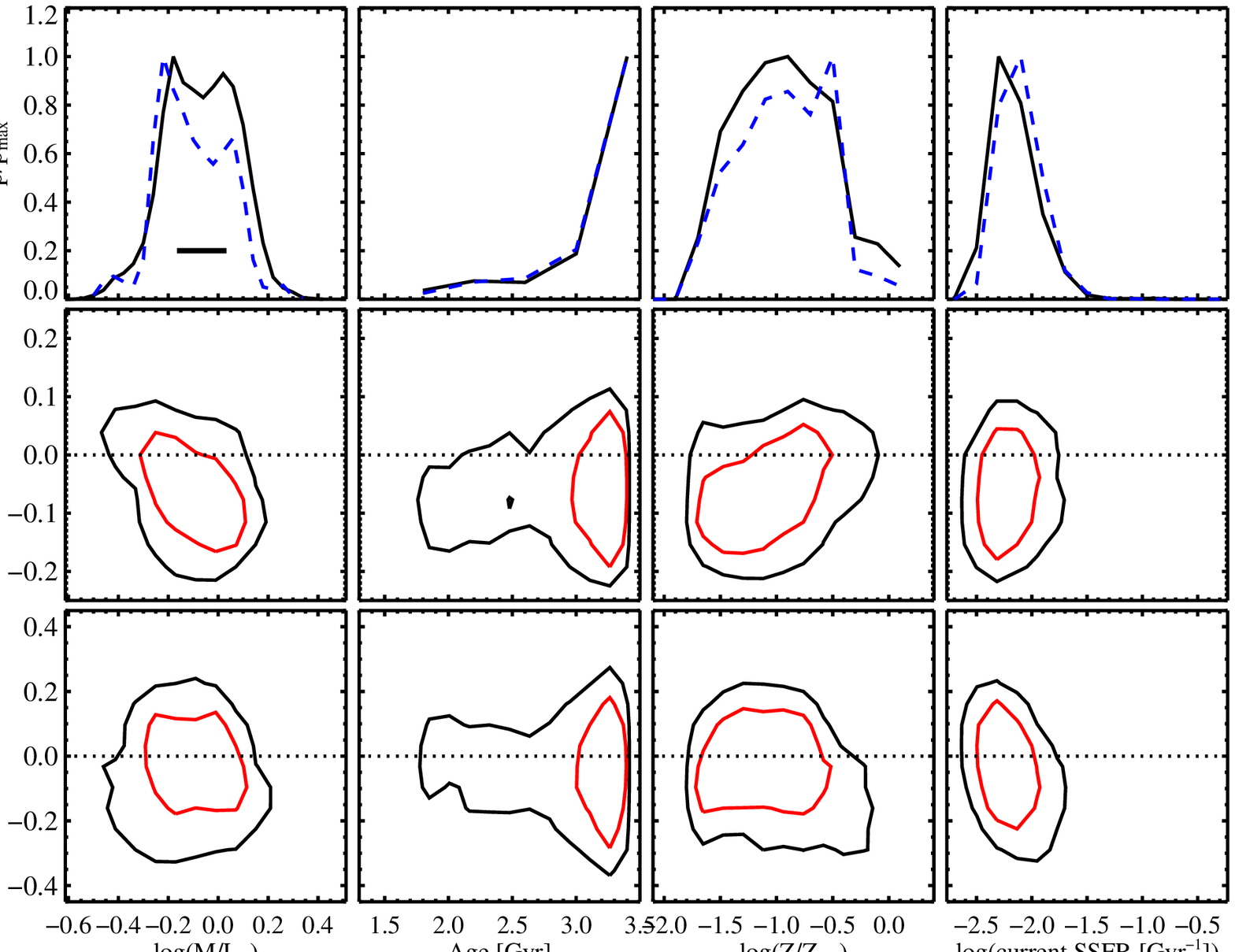}
\vspace{0.5cm}
\caption{Same as Figure \ref{fig:cgrid_bblu}, except now for a bright
  red galaxy at $z\sim2$.  The horizontal line in the upper-left panel
  indicates the 95\% CL limit on ${\rm log}(M/L_K)$ for this galaxy as
  reported by \citet[][this is galaxy 3650 according to their Table
  3]{Maraston06}, whose full width is 0.2 dex.  Accorinding to our SPS
  model, the 95\% CL limit on ${\rm log}(M/L_K)$ for this galaxy is
  0.6 dex.  The age of the Universe at the redshift of this object is
  3.6 Gyr for our adopted cosmology.  The likelihood contours
  involving the parameters $f_{BHB}$ and $S_{BS}$ are not included
  here because those aspects of stellar evolution are assumed to turn
  on in our model for ages $>5$ Gyr, which is older than the age of
  the Universe at the redshift of this galaxy.}
\label{fig:cgrid_ma06}
\vspace{0.5cm}
\end{figure*}

\subsection{Data at $z\sim2$}

In addition to data at $z\sim0$, we also analyze four
spectroscopically-confirmed galaxies at $z\sim2$ reported by
\citet{Daddi05}.  These galaxies lie in the {\it Hubble Ultra Deep
  Field} and have an elliptical morphology.  Spectral features and
broad-band colors indicate that the light from these galaxies is
dominated by A- or F-type stars, and are hence not currently
star-forming \citep{Daddi05, Maraston06}.  These galaxies are
interesting to study in detail because they constitute the data used
by \citet{Maraston06} to demonstrate the large systematic differences
between predictions from the SPS models of \citet{Maraston05} and
\citet{Bruzual03}.  According to \citet{Maraston06}, these galaxies
are very massive --- typical stellar masses are $\sim10^{11}\Msun$.

A wide array of broad-band photometry is available for these galaxies
including $VRJK$, {\it Spitzer IRAC} imaging in the $3.6\mu m$,
$4.5\mu m$, $5.8 \mu m$, and $8.0\mu m$ filters, and {\it Hubble Space
  Telescope} imaging in the F435, F606, F775, F895, F110, and F160W
filters.  \citet{Daddi05} originally presented results for seven
galaxies.  We use only those galaxies that are detected in all filters
and with $z>1.7$, leaving us with a sample of four.  Finally, we do
not use the $8.0 \mu m$ filter because our stellar libraries do not
extend that far into the infrared.

\subsection{Fitting procedure}
\label{s:fit}

The SPS model outlined in $\S$\ref{s:methods} and \ref{s:uncert}
contains 10 free parameters, summarized in Table \ref{t:models} (we do
not attempt to marginalize over the IMF parameter $m_c$).  Recall that
each point in this parameter space is associated with a spectrum of a
composite stellar population that can be transformed into a set of
broad-band colors.  Each set of parameters also specifies a
time-dependent mass-to-light ratio, $M/L$, in a variety of filters.
Our goal now is to explore the likelihood surface in this 11
dimensional space for each of a large number of observed galaxies.  In
particular, we are interested in the likelihood of $M/L$ for a given
galaxy.

Exploration of such a large parameter space is most efficiently
accomplished with a Monte Carlo Markov Chain (MCMC) algorithm.  In the
MCMC algorithm, a step is taken in parameter space and this step is
accepted if the new location has a lower $\chi^2$ compared to the
previous location, and is accepted with probability $e^{-\Delta
  \chi^2/2}$ if the $\chi^2$ is higher than the previous location.
Each step in the chain is recorded.  Since the acceptance of a step is
dependent on $\chi^2$, most of the time will be spent in regions of
high likelihood.  After a sufficient number of steps the likelihood
surface produced by the chain will converge to the true underlying
likelihood.  Convergence is defined according to the prescription
described in \citet{Dunkley05}.

A set of priors are specified in order to avoid un-physical regions of
parameter space.  The TP-AGB parameters, $\Delta_L$ and $\Delta_T$,
have Gaussian priors with zero mean and standard deviations of 0.15
and 0.10 respectively.  Uniform priors are adopted for all other
parameters over the ranges shown in Table \ref{t:models}.  All
quantities are logarithmically incremented except for the TP-AGB
parameters, which are already defined in the logarithm, and $T_{\rm
  start}$, which is incremented in the quantity ${\rm log}(T_{\rm
  univ}-T_{\rm start})$, where $T_{\rm univ}$ is the age of the
Universe at the redshift of the galaxy.  The ranges for the parameters
that incorporate uncertainties in stellar evolution are motivated in
$\S$\ref{s:uncert}.

\section{Results: from light to physical properties}
\label{s:res}

This section contains the results from fitting our SPS model to data
both at $z\sim0$ and $z\sim2$.  In this section we perform fits to the
data both for a `default' model where the isochrones are not modified
($\Delta_T=0$, $\Delta_L=0$, $f_{BHB}=0$, and $S_{BS}=0$) and
for a model that allows flexibility in uncertain phases of stellar
evolution, as discussed in detail in $\S$\ref{s:uncert}.  The goal
here is to investigate to what extent, if any, the uncertainties in
stellar evolution propagate into the central values and uncertainties
of physical properties such as stellar masses, mass-weighted ages, and
star formation rates of galaxies.

Here our aim is to highlight with representative examples the main
trends and results from fitting our SPS model to data.  To this end we
have chosen to focus on four classes of galaxies at $z\sim0$: bright
red, faint red, bright blue, and faint blue galaxies as defined in
Table \ref{t:sdss}.  

In addition, we discuss results from fitting our model to
passively-evolving luminous galaxies at $z\sim2$.  As mentioned above,
we have chosen this last set of galaxies because such galaxies were
studied by \citet{Maraston06} who found that the different treatments
of the TP-AGB phase in the SPS codes of \citet{Bruzual03} and
\citet{Maraston05} lead to large, systematic differences in stellar
masses and ages of galaxies.  The \citet{Maraston05} models yielded
galaxies that were both younger and less massive than the models of
\citet{Bruzual03}.  Since TP-AGB stars dominate the stellar light at
ages of $\sim2$ Gyr, we expect these passive galaxies at $z\sim2$,
which are dominated by stars at this age, to be the galaxy type most
sensitive to the uncertainties associated with the TP-AGB phase.  They
can thus be considered a `worst-case scenario' for the propagation of
uncertainties in stellar evolution into parameters such as stellar
mass and age.

Future work will focus on the consequences of the results presented in
this section with a more statistically complete sample of galaxies.

\subsection{Central values and uncertainties}

\begin{figure}[t!]
\plotone{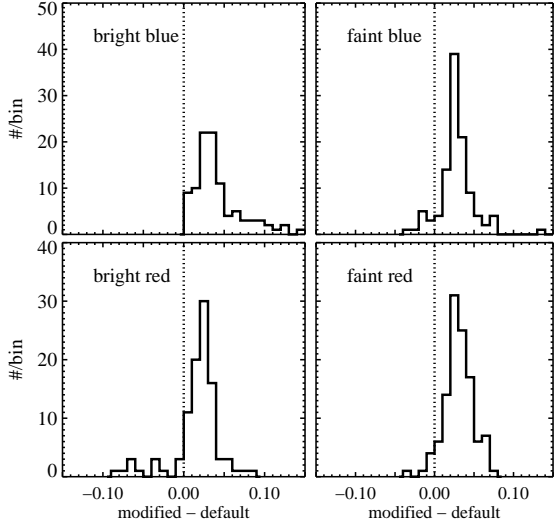}
\vspace{0.5cm}
\caption{Difference between the modified and default SPS models for
  the best-fit ${\rm log}(M/L_r)$ for galaxies of different types at
  $z\sim0$.  The modified models, which incorporate uncertain aspects
  of stellar evolution into the SPS formalism, favor slightly heavier
  galaxies, though the difference is well within the typical
  uncertainties ($\sim0.2-0.4$ dex).  Note that there is no systematic
  trend between the four different galaxy types defined in Table
  \ref{t:sdss}.}
\label{fig:mldiff}
\vspace{0.5cm}
\end{figure}

Figures \ref{fig:cgrid_bblu} and \ref{fig:cgrid_ma06} show results
from fitting our SPS model to the broad-band photometry of an $\sim
L^\ast$ blue galaxy at $z\sim0$, and a luminous red galaxy at
$z\sim2$, respectively.  The top row of panels show the marginalized
probability distribution for the mass-to-light ratio, mean age,
metallicity, and current specific star formation rate ($SSFR\equiv
SFR/M$).  Results are shown for both a default model where the
isochrones are fixed ({\it dashed lines}) and a model where aspects of
the isochrones are parameterized and marginalized over ({\it solid
  lines}).  The arrows in the first panel indicate the best-fit
$M/L_r$ from \citet{Kauffmann03a} and \citet{Blanton07}.

The lower four rows highlight the dependence of these four physical
parameters on four parameters that characterize uncertain aspects of
stellar evolution.  These parameters are the temperature and
luminosity of the TP-AGB phase ($\Delta_T$ and $\Delta_L$), the
fraction of blue horizontal branch stars ($f_{BHB}$), and the specific
frequency of blue straggler stars ($S_{BS}$).  Both 68\% and 95\% CL
contours are included.  The dotted lines indicate the values of the
parameters assumed in the default model.  In Figure
\ref{fig:cgrid_ma06}, plots associated with the parameters $f_{\rm
  BHB}$ and $S_{BS}$ have been omitted because the age of the
Universe at the redshift of this galaxy (3.6 Gyr) is younger than when
these parameters are assumed to turn on (i.e. for stellar ages older
than 5 Gyr).

\begin{figure}[t!]
\plotone{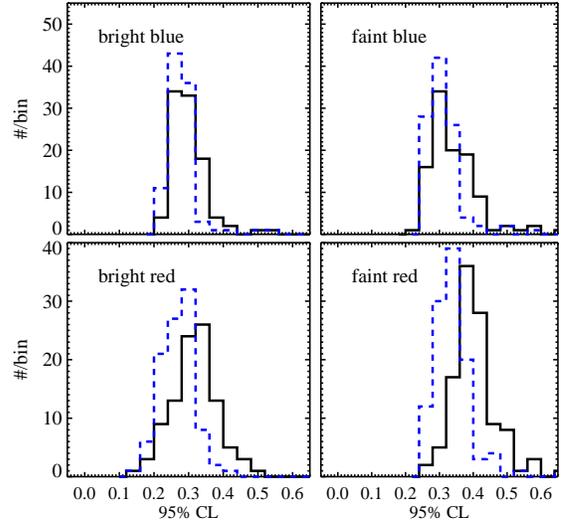}
\vspace{0.5cm}
\caption{95\% confidence limits on ${\rm log}(M/L_r)$ for samples of
  of galaxies at $z\sim0$ defined according to their $r-$band
  luminosities and $g-r$ colors.  For each galaxy, the 95\% CL is
  computed both for an SPS model that incorporates uncertainties in
  stellar evolution ({\it solid lines}) and a model that does not
  include such uncertainties ({\it dashed lines}). }
\label{fig:mlerr}
\vspace{0.5cm}
\end{figure}

The most striking result from these two figures is the lack of a
strong degeneracy between any of the physical parameters (along the
$x$-axis) and any of the parameters quantifying uncertain aspects of
stellar evolution (along the $y$-axis).  In Figure
\ref{fig:cgrid_bblu} there are weak degeneracies between TP-AGB
parameters ($\Delta_T$ and $\Delta_L$) and $SSFR$ and metallicity.
The lack of any strong degeneracies in Figure \ref{fig:cgrid_ma06} is
even more surprising in light of the fact that the \citet{Maraston06}
and \citet{Bruzual03} SPS models, which have different treatments of
the TP-AGB phase, produce systematically different stellar masses.  In
other words, these two SPS models have different values for $\Delta_T$
and $\Delta_L$, so one might have expected a degeneracy between $M/L$
and either or both of these TP-AGB parameters.  The reason for the
lack of any strong dependence in the present work is due to the
varying quality of fit for different TP-AGB parameters; this issue is
discussed in detail in $\S$\ref{s:marg}.

Despite a lack of strong degeneracies between parameters, the
marginalized probability distributions for the physical parameters are
clearly broader, in certain cases, for the modified compared to the
default models.  In particular, in Figure \ref{fig:cgrid_bblu} the
probability distributions of preferred metallicities and current
specific star formation rates are substantially broader for the
modified models.  This indicates that our knowledge of these
quantities is substantially impacted by uncertainties associated with
stellar evolution calculations.  Notice also that the uncertainty
associated with $M/L_K$ in Figure \ref{fig:cgrid_ma06} is a factor of
three larger than quoted in \citet{Maraston06}.  The reason for this
difference is not immediately clear, but may be due to the fact
that we allow for more flexibility in all aspects of the SPS modeling
including the SFH and dust prescription.

The parameters $f_{BHB}$ and $S_{BS}$ are essentially unconstrained by
the data (the likelihood contours span the entire range allowed by our
priors).  This is due to the combined facts that these parameters are
only important in the blue bands (one can see in Figures \ref{fig:bhb}
and \ref{fig:sbs} that they are of minor importance for bands redward
of $V$), and the bluest band included in the fits at $z=0$, the
$u$-band, carries a much larger uncertainty than the other bands (a
factor of $2-4$ times larger).  Better data in the blue and
ultraviolet could potentially yield tighter constraints, although the
impact of dust at these wavelengths will tend to weaken any
constraints.

\begin{figure*}[t!]
\plotone{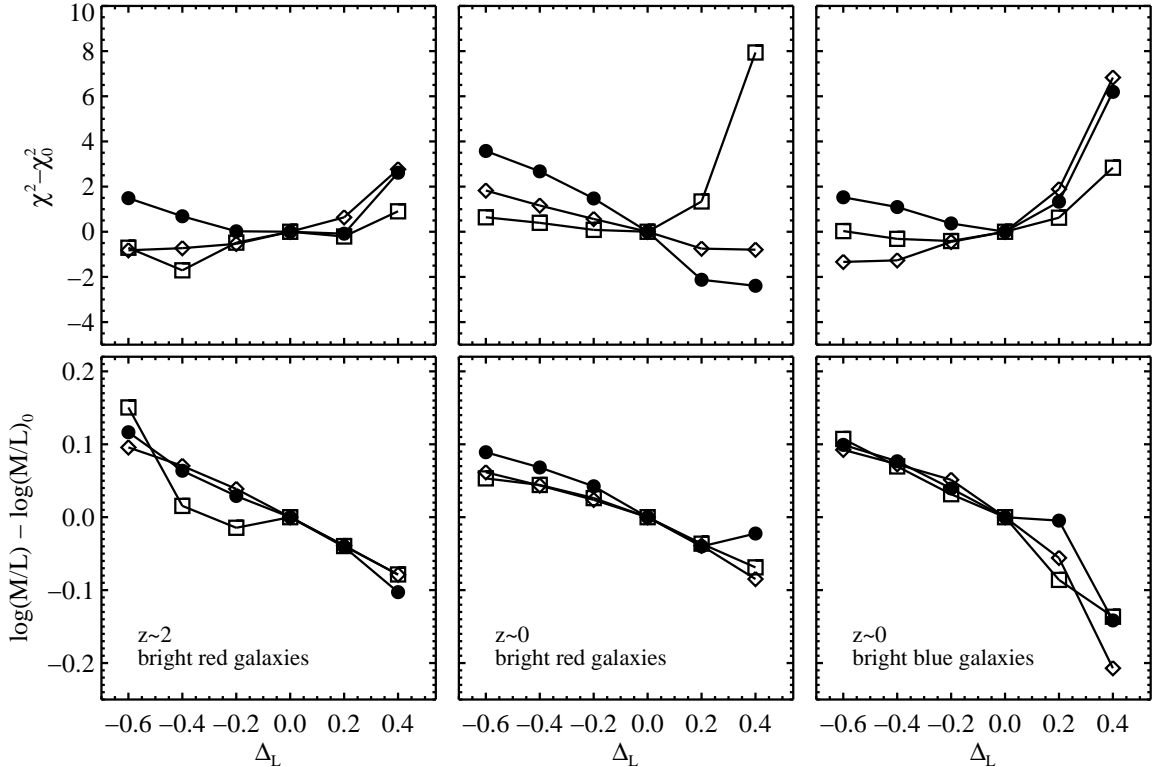}
\vspace{0.5cm}
\caption{Dependence of the best-fit $M/L$ and the minimum of $\chi^2$
  on the TP-AGB parameter $\Delta_L$ (see Table \ref{t:models} for a
  definition of this quantity).  We plot both the best-fit $M/L$ and
  the minimum of $\chi^2$ relative to the default model where
  $\Delta_L=0$ and $\Delta_T=0$.  In this figure we do not marginalize
  over the TP-AGB parameters but instead run models for fixed values
  of the parameter $\Delta_L$, while keeping $\Delta_T=0$.  Trends are
  shown for fits to three random galaxies ({\it solid, diamond, and
    square connected symbols}) from each of the following samples:
  bright red galaxies at $z\sim2$ (\emph{left column}), bright red
  galaxies at $z\sim0$ ({\it center column}), and bright blue galaxies
  at $z\sim0$ ({\it right column}). The trend of $M/L$ with $\Delta_L$
  implies that use of poor TP-AGB isochrones can systematically bias
  the resulting $M/L$ measurements and highlights the importance of
  marginalizing over the TP-AGB parameters, both at $z\sim2$ and
  $z\sim0$.}
\label{fig:fixmod}
\vspace{0.5cm}
\end{figure*}

\begin{figure*}[t!]
\plotone{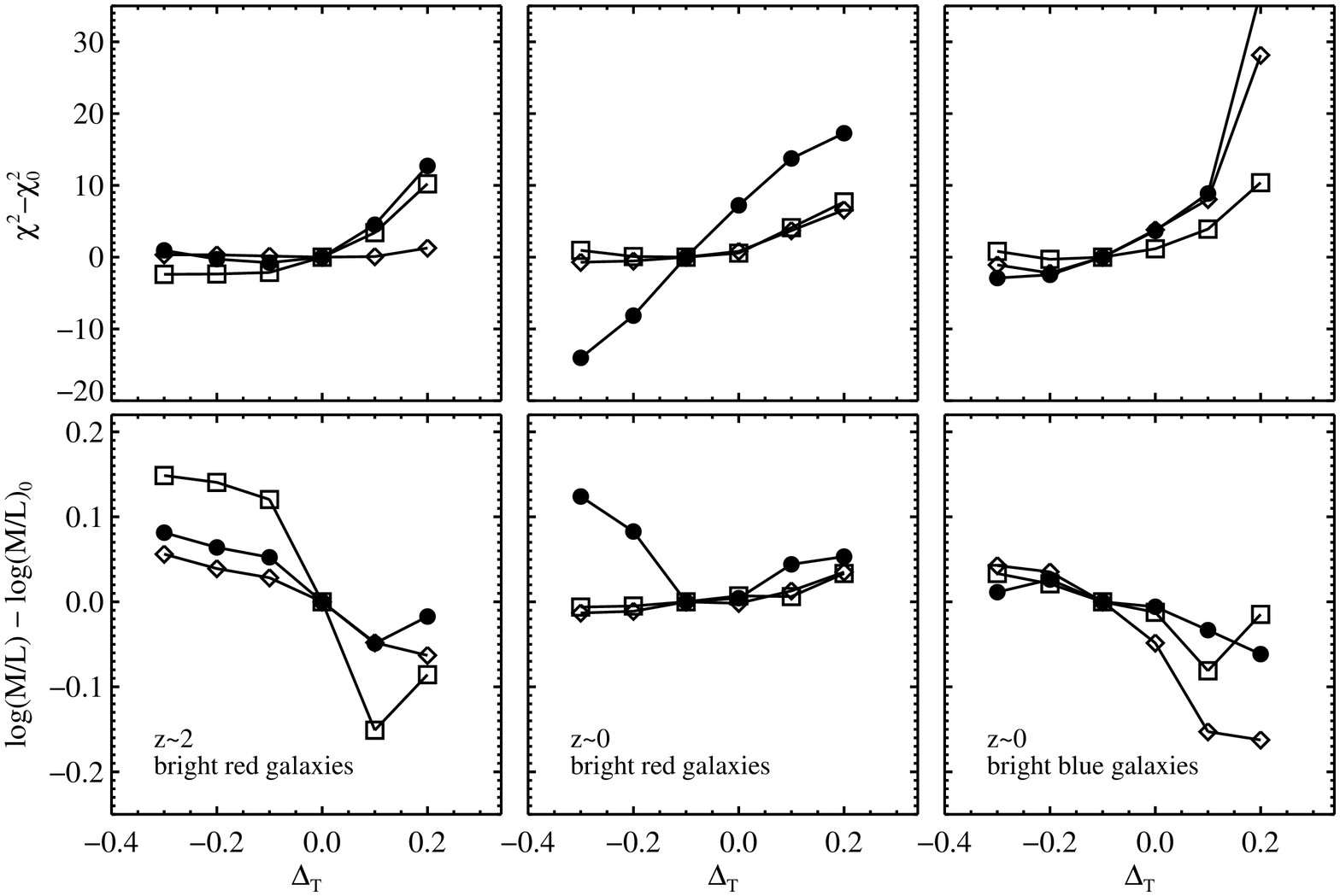}
\vspace{0.5cm}
\caption{Same as Figure \ref{fig:fixmod} except now plotting the
  best-fit $M/L$ and and the minimum of $\chi^2$ as a function of the
  TP-AGB parameter $\Delta_T$, while keeping $\Delta_L=0$.}
\label{fig:fixmod_delt}
\vspace{0.5cm}
\end{figure*}

Figure \ref{fig:mldiff} plots the difference between the modified and
default SPS models for the best-fit quantity ${\rm log}(M/L_r)$ for
the four sets of SDSS-2MASS galaxies described in $\S$\ref{s:data}.
There is a systematic offset for all galaxy types of $\sim0.05$ dex in
the sense that the modified models favor slightly higher stellar
masses.  The magnitude of this effect is small compared to the typical
errors (see the following paragraph), and there appears to be no
systematic trend with galaxy type.  These two results indicate that
the central values of $M/L_r$ (and hence of stellar mass), are not
particularly sensitive to the uncertain aspects of stellar evolution
that we have incorporated into the modified models with respect to the
particular default model that we have used.  Of course, use of a
different default model, such as the one used by \citet{Bruzual03},
would undoubtably lead to larger differences.

In Figure \ref{fig:mlerr} we show the estimated 95\% confidence limits
on ${\rm log}(M/L_r)$ for four types of galaxies at $z\sim0$.
Uncertainties are estimated using both the modified ({\it solid
  lines}) and default ({\it dashed lines}) SPS models.  The typical
uncertainties for blue galaxies is $\sim0.3$ dex and is similar for
both the modified and default models.  In contrast, the typical
uncertainties for the red galaxies are larger by $\sim0.05$ dex when
using the modified SPS model compared to the default model.  Moreover,
the uncertainties range from $\sim0.2$ to $>0.4$ dex, with typical
values of $\gtrsim0.3$ dex for the modified models.  

This result is surprising at first glance since red galaxies are
generally considered to have simpler star formation histories when
compared to blue galaxies.  The larger errors associated with red
galaxies is driven by the fact that the best SPS model fits are
poorer, as determined by the minimum of $\chi^2$, for the red compared
to blue galaxies.  This fact is likely related to the longstanding
problem of SPS models predicting colors redder than observations by as
much as $\sim0.1$ magnitudes \citep{Eisenstein01, Wake06}.  It is
intriguing that even with the increased flexibility of our SPS model
we are unable to obtain adequate fits to the red galaxy data (as
compared to the blue galaxies).  The resolution to this problem may
thus lie with other aspects of SPS modeling that we have not
explicitly investigated (see $\S$\ref{s:other} for a discussion of
other uncertain aspects of SPS modeling not explored herein).  Very
recently, \citet{Maraston09} has shown that the use of empirical
stellar spectra (as opposed to empirically-calibrated theoretical
spectra) results in a much better fit to the SDSS LRG colors.  This is
an intruiging result that clearly requires further attention.

At redshift $z\sim2$, all four luminous red galaxies studied herein
carry similar stellar mass uncertainties of $\approx0.6$ dex (95\% CL).

\subsection{The importance of marginalizing over uncertain phases of
  stellar evolution}
\label{s:marg}

In the previous section we found that the central values and
uncertainties in the physical properties of galaxies such as star
formation rates, stellar masses, ages, and metallicities, are not
strongly effected when incorporating uncertainties in stellar
evolution such as blue straggler stars, horizontal branch morphology,
and TP-AGB temperature and luminosity.  At first glance these
conclusions are surprising because of recent work that has shown that
different treatments of the TP-AGB phase lead to systematically
different estimates for the masses of galaxies \citep{Maraston06}.  In
this section we explore this apparent tension in detail.

In Figures \ref{fig:fixmod} and \ref{fig:fixmod_delt} we plot the
best-fit $M/L$ for a sample of galaxies at $z\sim2$ and at $z\sim0$ as
a function of the TP-AGB luminosity offset $\Delta_L$ and $\Delta_T$,
respectively.  Here we are not marginalizing over these parameters but
are instead running SPS models where we have fixed these parameters to
specific values.  Each SPS model will thus return a best-fit $M/L$ and
a corresponding minimum in $\chi^2$.  It is apparent from the figure
that $\Delta_L$ and $M/L$ are anti-correlated and so, to a lesser
extent, are $\Delta_T$ and $M/L$.  This is the trend one expects
because an observed amount of flux in the near-IR, where the TP-AGB
stars contribute a significant fraction of their power, can be fit by
either more stars at a lower luminosity or fewer stars at higher
luminosity.
 
This trend qualitatively echos the conclusions reached by
\citet{Maraston06} who found that more luminous TP-AGB stars lead to a
lower estimated stellar mass.  The difference between these
conclusions and the ones reached in the previous section, where it was
found that the TP-AGB parameters and $M/L$ were not substantially
degenerate, lies in the goodness-of-fit between model and data,
plotted in the upper panels of Figures \ref{fig:fixmod} and
\ref{fig:fixmod_delt}.  It is apparent from the figures that the
changes in $\chi^2$ are in most cases not significant, and thus the
data allow for a wide range in $\Delta_L$, $\Delta_T$ and $M/L$.  We
stress, however, that if one {\it fixes} the underconstrained TP-AGB
parameters (as one implicitly does when choosing a particular version
of a stellar evolution calculation) and performs SPS modeling of
observational data, then one can in principle introduce systematic
uncertainties as large as $\sim0.2$ dex into the best-fit $M/L$
values.

In short, use of poor stellar evolution models can have systematic
effects on the inferred physical properties of galaxies, such as
$M/L$, unless parameters are included that encapsulate the uncertain
aspects of stellar evolution directly into the fit between model and
data. 

In our approach we fit simultaneously for the physical properties of
galaxies and the uncertain phases of stellar evolution.  We can thus
ask if the default libraries adequately describe the data (i.e. if the
parameters capturing the uncertain phases are consistent with zero),
or if the current models are deficient in certain regimes.

\begin{figure}[t!]
\plotone{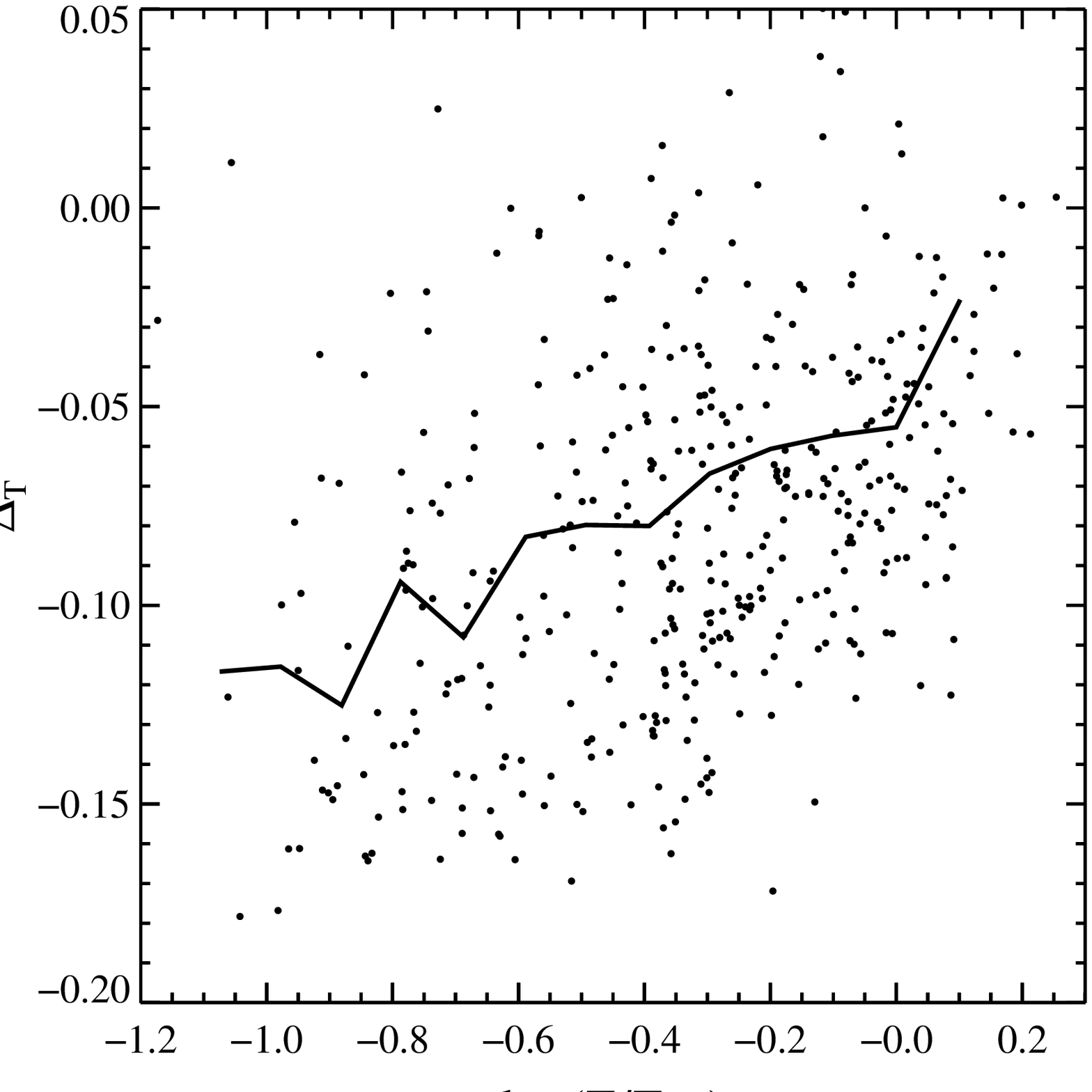}
\vspace{0.5cm}
\caption{Dependence of the best-fit logarithmic temperature offset for
  TP-AGB stars on the best-fit metallicity for all galaxies studied
  herein ({\it symbols}).  The median offset as a function of
  metallicity is also included ({\it solid line}).  The temperature
  offset for the TP-AGB stars is defined with respect to the current
  stellar models, which themselves depend on metallicity.  There is a
  clear trend that lower metallicity stars have systematically cooler
  TP-AGB stars than the current models predict.  This is not
  necessarily surprising because the stellar models (including both
  stellar evolution and stellar atmospheres) are not calibrated at low
  metallicity to the precision obtained in this figure.}
\label{fig:delt_zmet}
\vspace{0.5cm}
\end{figure}

In Figure \ref{fig:delt_zmet} we plot the best-fitting TP-AGB
temperature offset from the default models, $\Delta_T$, against the
best-fit metallicity for all galaxies studied herein.  It is clear
that near solar metallicity the current stellar models used herein
can adequately describe the broad-band SDSS+2MASS data.  This is not
surprising because the TP-AGB models have been calibrated largely on
stars in the solar neighborhood, where the mean metallicity is
approximately solar \citep{Marigo08}.

At sub-solar metallicities the temperature offset deviates
substantially from zero, on average, indicating that the default
models cannot adequately describe the data.  The offset is in the
sense that the data favor cooler TP-AGBs than the default stellar
models predict.  It is not altogether surprising that the default
models fail in this regime because they have not been adequately
tested at such low metallicities.  Aside from the Milky Way, the
TP-AGB isochrones have been calibrated against stars in the LMC and
SMC, which have metallicities $\sim1/3$ and $\sim1/5 \,Z_\Sol$,
respectively.  Perhaps more importantly, the empirical TP-AGB spectra
are largely from the solar neighborhood and both the metallicity and
the metallicity dependence are not well understood \citep{Lancon02}.

Another possible explanation is that this TP-AGB parameter is
compensating for another unrelated shortcoming of the current
generation of SPS models.  For example, as mentioned in
$\S$\ref{s:stellib}, the current stellar spectra libraries cannot
simultaneously fit both the observed globular cluster data and the
temperature-color relations of individual stars.  It is beyond the
scope of the present work to investigate this issue further.

These results suggest that current models are still inadequate in
untested regimes, and should thus be treated with caution where
extrapolations are required.  We find no other strong dependence
between physical parameters such as metallicity, star formation rate,
mass, or age on uncertain phases of stellar evolution.

\section{Discussion}
\label{s:disc}

\subsection{The importance of accurate uncertainties}

The importance of accurate and robust estimates for not only the
central values of various physical properties of galaxies but also
their associated uncertainties cannot be underestimated.  We highlight
here several areas where accurate uncertainties play an important, if
often neglected role.

With the advent of large galaxy surveys and publically-available SPS
codes, the estimation of stellar mass functions has become routine in
both the local and distant Universe \citep[see e.g.][]{Cole01, Bell03,
  Fontana04, Drory04, Bundy05, Borch06}.  Since the mass function
declines exponentially at the massive end, any uncertainty in the
masses of individual galaxies can lead to a substantial misestimation
of the underlying mass function.  \citet{Cattaneo08} provide
illustrative examples of the extent to which uncertainties can broaden
an intrinsically steep mass function.  This important effect is rarely
taken into account when either trying to estimate the underlying mass
function or when trying to compare to models of galaxy formation.
With accurate and reliable uncertainties, one can begin to deconvolve
the true mass function from the broadening effects of uncertainties.
This task is critical if one is to make accurate comparisons to models
of galaxy formation and evolution.

This concern does not reside solely with the mass function.  There are
many relations that are effected, including the relation between star
formation rate and stellar mass, the star formation rate function, and
the relation between stellar mass and large scale clustering strength.
This last relation is essential for understanding the connection
between galaxies and the underlying dark matter structure --- an
accurate accounting of the uncertainties is thus essential.

\subsection{Why the IMF matters}
\label{s:whyimf}

As discussed in $\S$\ref{s:uimf}, it is extremely difficult to
constrain the slope of the IMF in the solar neighborhood, let alone in
external galaxies.  This is unfortunate given the immense importance
of the IMF in many areas of astrophysics.

For example, the IMF is not often considered a source of uncertainty
when attempting to constrain the evolution of the luminosity function
of galaxies through time \citep[though see][who discuss this source of
uncertainty]{Cool08}.  In recent years it has become common to
construct the luminosity function for passively evolving (i.e. red)
galaxies at multiple epochs in order to understand how these galaxies
grow with time.  Recently, several authors have found evidence for
very little intrinsic evolution in the bright end of the luminosity
function of passive galaxies, suggesting that massive galaxies have
evolved little since $z\sim1$ \citep[e.g.][]{Wake06, Brown07, Cool08}.

Passive galaxies are considered relatively easy to study because, in
the absence of any appreciable star formation, the time-dependence of
the luminosity of such systems is determined simply by stellar
evolution and the IMF.  Authors take expected luminosity evolution
between two epochs for fiducial stellar evolution models and an IMF
and compare to the evolution in the luminosity functions between the
same two epochs in order to infer the underlying (intrinsic) change in
the population with time.

We found in $\S$\ref{s:uimf} however, that the uncertainty in the
slope of the IMF in the solar neighborhood translates into an
uncertainty in the amount of passive luminosity evolution of $\sim0.4$
magnitudes per unit redshift in the $K-$band.  The difficulty of
accounting for systematics in the observed IMF, and the unique
difficulty of measuring the IMF slope near $1\Msun$ --- precisely
where it must be measured accurately for passive evolution corrections
--- suggest that this uncertainty in luminosity evolution could be an
underestimate.  Any possible evolution in the IMF, as discussed in
$\S$\ref{s:uimf}, will only exacerbate this problem.

This uncertainty can have serious consequences for understanding the
evolution of galaxies.  We highlight one example here.  There has been
some tension in the literature between merger rates of luminous
passive systems inferred from close-pair counts and from the evolution
of the luminosity function.  Taken at face value, constraints from the
luminosity function suggest that the merger rate amongst bright,
passive galaxies has been very modest since $z=1$ \citep{Wake06,
  Brown07, Cool08}, in contrast to the apparently higher rates
inferred from the close pair counts \citep{vanDokkum05, Bell06b,
  Masjedi06, Lin08}.

If the slope of the IMF were shallower than the canonical value, then
from Figure \ref{fig:lumev} we conclude that luminosity evolution
would be stronger than is typically assumed.  If this were the case,
then evolution of the luminosity function would instead require a
higher merger rate than inferred with the canonical IMF slope. The
slope of the IMF is thus intimately related to the merger rate as
inferred from evolution of the luminosity function.

A comprehensive comparison between various merger rate indicators has
yet to be undertaken, and the apparent tension in the literature
between indicators may be largely interpretive.  If however there is a
quantitative tension, a non-canonical IMF slope may reconcile
different measures.

An additional and important constraint is the evolution of the
fundamental plane.  As discussed in detail in \citet{vanDokkum08}, the
evolution in the fundamental plane can provide constraints on the
logarithmic slope of the IMF because one can measure and compare mass
evolution to luminosity evolution, so long as one is confident that
the descendants of high redshift galaxies can be reliably identified
at low redshift.  There are additional assumptions that make it
difficult in practice to set strong constraints.  For example, this
approach requires knowledge of the fraction of mass that is attributed
to dark matter.  Nonetheless, use of additional constraints, such as
the fundamental plane and its evolution, can provide valuable
additional constraints not only on the form of the IMF but also on
uncertain aspects of stellar evolution.

\subsection{Spectroscopy and the SDSS}
\label{s:spec}

We have focused on fitting our SPS model to the broad-band photometry
of galaxies in the near-UV through near-IR.  We now discuss the use of
observed spectra in constraining SPS models.

In addition to broad-band photometry, the SDSS provides reasonable
signal-to-noise, medium resolution ($R\sim2000$) spectroscopy for
$\sim670,000$ galaxies as of DR4.  These spectra provide a substantial
amount of information not available in broad-band photometry.  Indeed,
several independent groups have made extensive use of spectroscopy to
infer physical properties of SDSS galaxies \citep[e.g.][]{Kauffmann03a,
  Panter07}.

There are several concerns related to using spectroscopy for SPS
modeling that have not received extensive attention in the SPS
literature.  First, the robustness of the stellar spectral libraries
is not known over the full luminosity, temperature, and metallicity
range required.  The TP-AGB spectra are clearly deficient, as
described in $\S$\ref{s:tpagb}, and will thus introduce uncertainty in
the interpretation of optical and near-IR spectral features.  As
discussed in $\S$\ref{s:stellib}, there are also known systematic
problems with the color-temperature relations in the stellar
libraries, indicating again that the libraries probably contain
unknown systematics.  Furthermore, the recent results of
\citet{Maraston09} suggest that the theoretical spectral libraries are
deficient in important ways.

An example of these complications was highlighted by \citet{Wild07},
who recently showed that there is a significant offset in spectral
indices between the BC03 model and SDSS data.  The authors attribute
the mis-match to the stellar spectral libraries used in the BC03 model
and caution that systematic errors in SPS models may be significantly
larger than quoted statistical errors.

A second source of concern, at least for SDSS galaxies, is that the
spectra are taken through fibers that only cover a fraction of the
total galaxy.  Averaged over the full spectroscopic sample with
$0.01<z<0.15$, the spectra only captures the light from $\approx30$\%
of the total galaxy.  For example, a galaxy may have both a bulge and
a disk but the fiber would only cover the bulge and thus the spectra
of this galaxy would not be representative of the total galaxy.

The standard procedure to deal with this fiber effect is to use the
observed spectra in the SPS fitting and then make a correction based
on the observed color(s) outside of the fiber \citep[see
e.g.][]{Brinchmann04}.  Given the unknown systematics in the spectral
libraries, in conjunction with the simple correction that is typically
adopted to account for light outside of the fiber, we do not believe
that global physical properties of galaxies derived from spectra are
more robust than information derived solely from broad-band photometry.

In fact, there is some evidence that the masses derived from spectra
contain significant systematic uncertainties.  \citet{Maller09} showed
that the stellar masses derived with spectroscopic information by
\citet{Kauffmann03a} are a function of the inclination of the galaxy
in the sense that face-on galaxies are on average $\approx 0.2$ dex
heavier than edge-on galaxies.  Of course, an intrinsic property of a
galaxy such as its stellar mass should not be a function of its
orientation projected on the sky.  This trend thus suggests a
systematic bias in the masses derived by \citet{Kauffmann03a}.  The
stellar masses estimated from broad-band photometry by
\citet{Blanton07} show a weaker dependence on inclination, and the
estimates based on broad-band photometry from \citet{Bell03} show no
trend with inclination.  It is not known if the trend in the
spectroscopically-derived masses is an intrinsic problem with using
spectroscopic data or a systematic in the methodology of
\citet{Kauffmann03a}.

\subsection{Additional uncertainties not explored herein}
\label{s:other}

There are additional uncertainties not explored herein that may
significantly impact the results of SPS modeling.  In this section we
briefly discuss some of the more important uncertainties.

In $\S$\ref{s:stellib} we discussed several of the limitations of the
current generation of stellar spectral libraries.  \citet{Martins07}
has presented a detailed comparison between several theoretical and
empirical libraries and found substantial disagreement for both blue
and red colors (such as $U-B$ and $V-K$, respectively).  The
difficulty in modeling stellar spectra is exacerbated by the fact that
empirical libraries span a relatively narrow range in metallicity.
Without a well-sampled empirical library it is thus difficult to
understand and isolate the limitations of the models \citep[see
also][]{Charlot96a, Charlot96b, Yi03}.

The impact of non-solar abundance ratios has not been discussed in the
present work.  Observations consistently show that the $\alpha$-to-Fe
ratio is a function of galaxy mass, with more massive galaxies being
more highly $\alpha$-enhanced \citep[e.g.][]{Thomas05}.  Despite this
well-known observational fact, all SPS models that attempt to
determine physical properties of galaxies from either their colors or
spectral shapes rely on stellar spectra and evolutionary calculations
with solar abundance patterns\footnote{A considerable amount of work
  has gone into re-calibrating the basic outputs of solar-scaled SPS
  models (the SSPs) so that they may be applied more generally to
  variable abundance ratios.  Such efforts attempt to use a variety of
  absorption line strengths as a diagnostic to determine the abudance
  patterns of individual galaxies.  This method is calibrated on the
  abundance patterns of globular clusters for which detailed abundance
  patterns can be measured on a star-by-star basis.  For further
  information see e.g. \citet{Worthey94, Thomas03}.}.  The effect of
this omission on derived physical properties of galaxies has not been
adequately quantified, but various calculations suggest that the
effect can be substantial, especially for spectral signatures
\citep[see e.g.][]{Coelho07}.  The complication here is two-fold
because the abundance patterns effect not only the spectra of
individual stars but also the isochrones and their evolution.  Fully
self-consistent models of $\alpha$-enhanced abundance patterns are
only just beginning \citep{Coelho07, Dotter07, LeeHC09}, but early
results suggest that the magnitude of this effect on colors can be as
high as 0.1 magnitudes \citep{Coelho07}.

We have not explored uncertainties in either the main sequence
turn-off point or the red giant branch.  \citet{Gallart05} has
presented an extensive comparison between the most popular stellar
evolution models and found that slightly different treatments in the
underlying physics can lead to detectable differences in derived
properties such as ages and metallicities \citep[see
also][]{Maraston05}.  For example, current stellar models predict
different times for the onset of the red giant branch, owing in part
to different treatments of convection in the interiors
(i.e. assumptions relating to convective core overshoot and to the
mixing length parameter).  These differences are large enough to be
detected in star clusters in the LMC \citep{Ferraro04}, and can impact
the broad-band colors of galaxies \citep{Yi03, LeeHC07}. 

Another source of uncertainty that is currently not treated in SPS
models is contamination of the galactic light from active galactic
nuclei (AGN).  Overall, the fraction of galactic light attributed to
AGN is small \citep[e.g.][]{Hao05}, but there are certain regimes
where the contribution may be non-negligible.  The problem can be
particularly acute when spectroscopic information is not available to
isolate AGN candidates.

\section{Summary}
\label{s:conc}

In this paper we have presented a novel SPS model that is capable of
flexibly handling various uncertain aspects of stellar evolution
including TP-AGB stars, the horizontal branch, blue stragglers, and
the IMF.  This model thus allows one to understand the relevance of
these uncertain aspects to the derived physical properties of
galaxies.  In particular, one can marginalize over these uncertain
aspects in order to understand the full uncertainties associated with
the physical properties of galaxies including stellar masses, mean
ages, metallicities, and star formation rates.  We have applied this
model to fit broad-band near-UV through near-IR photometry of a
representative sample of galaxies at $z\sim0$ and $z\sim2$.  We have
also explored the effect of the IMF on the observable properties of
galaxies and have compared the stellar populations of single- to
multi-metallicity stellar systems.

Significant results include the following:

\begin{itemize}

\item[1.]

  When including the uncertainties in various stages of stellar
  evolution, stellar masses determined from broad-band near-UV through
  near-IR photometry at $z\sim0$ carry uncertainties of at least
  $\sim0.3$ dex at 95\% CL.  The masses of luminous red galaxies at
  $z\sim2$ are uncertain at the $\sim0.6$ dex level.  The
  uncertainties in stellar mass are not strongly correlated with
  galaxy color or luminosity.

\item[2.]

  The TP-AGB phase in stellar evolution is clearly important for
  understanding the physical properties of galaxies.  For example, an
  inadequate treatment of this uncertain phase will lead to
  substantial and {\it systematic} mis-estimation of stellar masses.
  Our SPS model suggests that either current stellar evolution
  calculations, model atmospheres, or both, do not adequately describe
  the metallicity-dependence of the TP-AGB phase.  In addition, there
  are various degeneracies between the temperature and luminosity
  scale of the TP-AGB phase on the one hand, and the inferred
  metallicity, stellar mass, and star formation rate, on the other
  hand.  Thus, models that do not include uncertainties in the TP-AGB
  phase are underestimating these derived physical properties and are
  potentially introducing systematic biases.

\item[3.]

  The uncertainties in the morphology of the horizontal branch and the
  frequency of blue straggler stars does not appreciably impact the
  SPS analysis.  This is due to the fact that these aspects become
  increasingly important in the ultraviolet, where the data used in
  this analysis carry the largest uncertainties.  Inclusion of more
  accurate ultraviolet data would probably highlight the importance of
  these phases in the total error budget.

\item[4.]

  The uncertainty in the logarithmic slope of the IMF, as determined
  for the solar neighborhood, implies an uncertainty in the
  luminosity-evolution of a passively evolving system of $\sim0.4$
  magnitudes per unit redshift in the $K-$band.  This is a substantial
  source of uncertainty that is rarely accounted for in discussions of
  the cosmic evolution of galaxy populations.  The uncertainties
  associated with luminosity evolution may be even more severe both
  because the logarithmic slope of the IMF has not been directly
  measured in any external galaxy and a redshift-dependent IMF cannot
  be ruled out by current data.

\item[5.]

  The broad-band evolution of a multi-metallicity population of stars
  is essentially equivalent to a single-metallicity population of
  stars whose metallicity is the mean of the multi-metallicity
  population, for bands redward of $V$.  This statement holds when one
  treats the morphology of the horizontal branch separately, as we
  have done herein.  As one considers bands bluer than $V$, systematic
  differences arise between single- and multi-metallicity populations
  that may cause systematic biases in quantities such as galaxy ages,
  star formation rates, and metallicities.

\end{itemize}

In this work we have focused our attention on a limited number of
broad--band filters.  If significant progress is made in understanding
the uncertainties in existing spectral libraries, it is possible that
spectroscopy may provide much stronger constraints on the quantities
of interest.  In addition to spectroscopy, we leave open the
possibility that carefully chosen broad-- and narrow--band filters may
provide stronger constraints than the filters explored herein.

For many of the uncertainties considered herein, we have adopted prior
ranges that are at the extremes of, or larger than, that suggested by
observational results.  This is particularly the case for the range of
metallicity distribution functions, horizontal branch morphology, and
blue straggler specific frequency.  We have demonstrated that
broad-band optical through near-IR photometry is not sensitive to
these uncertainties with our adopted priors, and therefore even a
pessimistic assessment of our knowledge of these aspects has little
effect on the derived properites of galaxies.  The uncertainties
listed above are more important in the ultraviolet, and we therefore
expect the assumed uncertainties and priors to play a more significant
role in the interpretation of restframe ultraviolet data.  Exploration
of these issues will be the subject of future work.

\acknowledgments 

We thank the stars for shining bright, and Tom Brown for providing his
globular cluster data.  This work has benefited substantially from
conversations with others.  We thank Raul Jimenez, Claudia Maraston,
Jerry Ostriker and Pieter van Dokkum for their opinions and insights,
and the referee for a careful report.  This project began at the Aspen
Center for Physics (partially funded by NSF-0602228) in the summer of
2007.  We thank the Center for their hospitality.

Funding for the Sloan Digital Sky Survey (SDSS) has been provided by
the Alfred P. Sloan Foundation, the Participating Institutions, the
National Aeronautics and Space Administration, the National Science
Foundation, the U.S. Department of Energy, the Japanese
Monbukagakusho, and the Max Planck Society. The SDSS Web site is
http://www.sdss.org/.

The SDSS is managed by the Astrophysical Research Consortium (ARC) for
the Participating Institutions. The Participating Institutions are The
University of Chicago, Fermilab, the Institute for Advanced Study, the
Japan Participation Group, The Johns Hopkins University, Los Alamos
National Laboratory, the Max-Planck-Institute for Astronomy (MPIA),
the Max-Planck-Institute for Astrophysics (MPA), New Mexico State
University, University of Pittsburgh, Princeton University, the United
States Naval Observatory, and the University of Washington.

This publication makes use of data products from the Two Micron All
Sky Survey, which is a joint project of the University of
Massachusetts and the Infrared Processing and Analysis
Center/California Institute of Technology, funded by the National
Aeronautics and Space Administration and the National Science
Foundation.

This work made extensive use of the NASA Astrophysics Data System and
of the {\tt astro-ph} preprint archive at {\tt arXiv.org}.

\bibliography{../master_refs}

\end{document}